\documentclass[aps,physrev,reprint,superscriptaddress,longbibliography,floatfix]{revtex4-2}
\usepackage{verbatim}
\usepackage{dsfont}
\usepackage{amsmath}
\usepackage{amsthm}
\usepackage{enumerate}
\usepackage{graphicx}
\usepackage{mathtools}
\usepackage{mathrsfs}
\usepackage{bm}
\usepackage{booktabs}
\usepackage{amssymb}
\usepackage{xcolor}
\usepackage{hyperref}
\usepackage{overpic}
\usepackage{placeins}
\usepackage{orcidlink}  
\usepackage[capitalize,nameinlink]{cleveref}
\hypersetup{
    colorlinks=true,
    linkcolor=blue,
    citecolor=blue,
    unicode=true,
}

\newcommand{\ket}[1]{\left| #1 \right\rangle}
\newcommand{\bra}[1]{\left\langle #1 \right|}
\newcommand{\braket}[2]{\left\langle#1\middle|#2\right\rangle}
\newcommand{\ketbra}[2]{\left|#1 \rangle \langle #2 \right|}

\DeclareMathOperator{\tr}{Tr}

\usepackage{calc}
\begin{document}

\title{High-fidelity quantum gate learning via physics-informed neural networks}

\author{Luis Fabra}
\email{lfabramartinez91@correo.unicordoba.edu.co}
\affiliation{Department of Physics and Electronics, University of C\'ordoba, 230002 Monter\'ia, Colombia}
\author{Cristian E. Susa\orcidlink{0000-0003-4178-6102}}
\email{cristiansusa@correo.unicordoba.edu.co}
\affiliation{Department of Physics and Electronics, University of C\'ordoba, 230002 Monter\'ia, Colombia}
\author{Alcides Montoya}
\affiliation{Centro de Excelencia en Computación Cuántica e Inteligencia Artificial, Universidad Nacional de Colombia, Sede Medellín, Colombia}
\author{Ariel Norambuena\orcidlink{0000-0001-9496-8765}}
\email{ariel.norambuena@usm.cl}
\affiliation{Departamento de Física, Universidad Técnica Federico Santa María, Casilla 110 V, Valparaíso, Chile}

\begin{abstract}
We propose a physics-informed neural network (PINN) approach for high-fidelity quantum gate learning in both one- and two-qubit systems. We train a PINN using the time-dependent Schr\"{o}dinger equation, where smooth control fields with fixed boundary conditions are obtained by minimizing a loss function that considers model dynamics, gate error, physical constraints, regularization terms, and multiple initial conditions. Training considers pairs of mutually unbiased bases, while independent validation combines Haar-random input states with the average gate fidelity. Unitary simulations yield an average gate infidelity of $8.49\times10^{-8}$ for Hadamard and values between $3.39\times10^{-7}$ and $4.96\times10^{-6}$ for controlled-NOT under $XX$, $YY$, and $ZZ$ interactions. We examine how the learned control fields depend on the coupling, training configuration, and available local fields, including five exchange-type dipolar scenarios. We test the learned control fields under the Lindblad master equation with dephasing and amplitude damping, theoretically reproducing the expected first-order fidelity loss in the weak-dissipation regime. Finally, boundary-adapted Chebyshev expansions represent 139 learned controls with a minimum common count of $K_\ast=24$ coefficients per control for all eleven family-averaged mean squared errors to fall below $10^{-6}$. Separately, square-pulse and Strang propagation at $K=30$ preserve all 31 continuous polynomial references within infidelity $10^{-5}$ at $N=128$ and all larger tested resolutions, under both sampling rules. These results connect multi-state gate synthesis, control-resource characterization, and compact square-pulse playback.
\end{abstract}

\maketitle

\section{Introduction}
\label{sec:intro}

Performing high-fidelity quantum gates is a critical challenge in developing practical, fault-tolerant quantum computers. Single- and two-qubit gates are the fundamental building blocks of quantum circuits, and their fidelity directly impacts the viability of quantum algorithms. Achieving high gate fidelities requires careful control-pulse optimization, motivating methods for optimal gate synthesis and numerical control design \cite{Niu2019,Chen2025}. The fundamental obstacle is navigating the high-dimensional control landscape while satisfying multiple objectives: maximizing gate fidelity, minimizing gate time, and ensuring robustness to experimental noise and parameter uncertainties. In most quantum platforms, including superconducting circuits, ion traps, and neutral atom systems, quantum gates are generated through carefully engineered time-dependent control Hamiltonians.

Designing these control functions remains a central challenge in quantum technologies because they must satisfy diverse physical constraints, including bounded amplitudes, limited bandwidth, robustness against decoherence, and resilience to experimental imperfections. Traditional quantum optimal control methods, such as GRAPE and CRAB \cite{Khaneja2005,Caneva2011}, have achieved remarkable success; however, they typically require repeated numerical optimization for each target gate and experimental configuration, which becomes computationally expensive and difficult to scale \cite{Riaz2019}. In this context, physics-informed neural networks (PINNs) offer a promising alternative by embedding the governing quantum dynamics directly into the training process, allowing neural networks to parametrize continuous state trajectories and control functions \cite{Raissi2019,Karniadakis2021}.

PINNs were originally introduced as a framework to solve forward, inverse, and nonlinear problems governed by partial differential equations by incorporating physical laws into the loss function of deep neural networks \cite{Raissi2019}. Since then, researchers have successfully applied PINNs across a broad spectrum of scientific disciplines, including fluid mechanics, materials science, engineering, and biology \cite{Cai2021,HAGHIGHAT2021,Karniadakis2021,GUO2025}. The key innovation of PINNs is their ability to encode physical laws directly into the loss function during neural network training, constraining the network to learn solutions consistent with underlying physics rather than relying solely on data fitting. Over the past few years, neural network-based approaches have demonstrated strong potential in quantum control, where machine learning models are trained to generate control pulses that steer quantum systems toward desired states or unitary operations~\cite{GUNGORDU2022,Lauten2025,Norambuena2024,Fosel2018}. In particular, Ref.~\cite{GUNGORDU2022} introduced a PINN-based scheme for obtaining smooth, noise-robust control pulses for unitary dynamics by directly minimizing sensitivity to quasistatic noise, without averaging over noise realizations. Reference~\cite{Lauten2025} demonstrated PINN-based two-qubit gate design for an isotropic Heisenberg model using process-fidelity objectives evaluated from propagated quantum channels under Schr\"{o}dinger and Lindblad dynamics.

In this work, we build on the framework of Ref.~\cite{Norambuena2024} to formulate gate synthesis through simultaneous density-matrix trajectories driven by shared control fields. Target transformations are imposed on states drawn from pairs of mutually unbiased bases, without requiring explicit construction of the full propagator in the training loss. We assess the resulting gates independently through numerical propagation, Haar-distributed test states, and average gate fidelity. We use this formulation to systematically investigate how the learned pulses depend on the interaction, training configuration, and available local controls, considering individual $XX$, $YY$, and $ZZ$ couplings and five exchange-type $XY$ configurations. An analytical ansatz enforces vanishing pulse amplitudes and first derivatives at both temporal endpoints. We further characterize dissipative fidelity losses using established weak-dissipation theory~\cite{abad2022universal,abad2025impact} and quantify compact waveform representations through boundary-adapted Chebyshev expansions. Together, these analyses connect multi-state gate synthesis with control-resource characterization, independent validation, and explicit pulse representation.

\section{Quantum system and logical gates}
\label{sec:sys_gate}

In this section, we introduce the physical models used to train the PINN and the target quantum logic gates. In particular, we introduce the problem of implementing the Hadamard and Controlled-NOT gates in terms of the unitary dynamics encoded in the time-dependent Schr\"{o}dinger equation, where control fields are the target functions to realize the quantum control task.

\subsection{Theoretical model for one- and two-qubit quantum dynamics}

In quantum mechanics, the dynamics of isolated systems is described by a unitary evolution. For a vector state in a $d$-dimensional Hilbert space, $\ket{\psi(t)} \in \mathcal{H}^{(d)}$, the Schr\"odinger equation reads ($\hbar = 1$)
\begin{equation} \label{eq:unitary_u}
\frac{d}{dt}|\psi(t)\rangle = -i H(t)|\psi(t)\rangle,
\end{equation}
or, equivalently, in terms of the unitary propagator $U(t)$ which satisfies $\dot{U}(t) = -i H(t)U(t)$ with the initial condition $U(0)=\mathds{1}$, where the Hamiltonian $H(t)=H_0 + \sum_{\alpha}H_{\alpha}(t)$ is usually split into a bare (time-independent) and a time-dependent part. The dynamics of a single qubit can be described by using the Pauli operators $\boldsymbol{\sigma} =(\sigma_x, \sigma_y, \sigma_z)$, which, together with the identity operator, form a complete operator basis on $\mathcal{H}^{(2)}$. In this representation, the Hamiltonian of a driven single-qubit system generally reads
\begin{equation}
H_{1}(t) = \boldsymbol{u}(t) \cdot \boldsymbol{\sigma},
\label{eq:SQH_gen}
\end{equation}
where $\boldsymbol{u}(t)=(u_x(t), u_y(t),u_z(t)) \in \mathbb{R}^3$ is a time-dependent control vector describing, for instance, external fields applied along the corresponding Bloch-sphere directions. These controls are typically constrained by physical limitations such as bounded amplitude, finite bandwidth, and smoothness imposed by the experimental devices.

Within this framework, arbitrary single-qubit gates can be synthesized by properly shaping the
control vector $\boldsymbol{u}(t)$. For the sake of implementation of our PINN scheme, we consider a system with a bare Hamiltonian along $\sigma_z$ and a transverse control Hamiltonian, such that \cref{eq:SQH_gen} reduces to~\cite{Fernandes_2023}

\begin{equation}
H_s(t) =  -\Delta \sigma_z + u_x(t)\sigma_x,
\label{eq:SQH}
\end{equation}
where $\Delta$ sets the static energy scale of the system, and $u_x(t)$ is the unknown control function. 

For a two-qubit system, the Hilbert space is $\mathcal{H}^{(4)}=\mathcal{H}^{(2)}\otimes\mathcal{H}^{(2)}$, and the
natural operator basis is given by tensor products of Pauli matrices
$\sigma_\alpha \otimes \sigma_\beta$, with $\alpha,\beta \in \{0,x,y,z\}$ and $\sigma_0=\mathds{1}$. We consider a Hamiltonian with local controls and diagonal qubit-qubit couplings,
\begin{align}
H_{2}(t) &= \sum_{\alpha }J_{\alpha}\, \left(\sigma_{\alpha} \otimes \sigma_{\alpha} \right) +  \sum_{j=1,2}\boldsymbol{u}^{(j)}(t) \cdot \boldsymbol{\sigma}^{(j)}   ,
\label{eq:tq_H}
\end{align}

where $\alpha \in\{x,y,z\}$, $\boldsymbol{u}^{(i)}(t)$ are the local control vectors acting independently on each qubit, and $J_{\alpha}$ denote the coupling strengths along the three Pauli directions. Here, $\boldsymbol{u}^{(1)}(t) \cdot \boldsymbol{\sigma}^{(1)} \equiv \sum_l u^{(1)}_{l}\sigma_l \otimes \mathds{1}$ and $\boldsymbol{u}^{(2)}(t) \cdot \boldsymbol{\sigma}^{(2)} \equiv \mathds{1} \otimes \sum_l u^{(2)}_{l}\sigma_l$, with $l \in\{x,y,z\}$. The above Hamiltonian is general enough to describe a wide variety of physical platforms, including superconducting qubits with tunable couplers, trapped ions with tunable, phonon-mediated spin–spin interactions, and neutral atom systems interacting via Rydberg blockade mechanisms. The coupling terms enable entanglement generation, which is essential for implementing two-qubit gates such as the controlled-NOT (CX) and controlled-Z gates.

For the systematic analysis presented in this work, we apply the PINN scheme to the Hamiltonian given in Eq.~\eqref{eq:tq_H} by considering each coupling term separately. Thus, the Hamiltonian to be used in our simulations with two qubits is of the general form
\begin{align}
H_{k\otimes k}(t)
&= J_k \, \sigma_k \otimes \sigma_k \nonumber\\
&\quad + \sum_{l} \left(
u_l^{(1)}(t)\, \sigma_l \otimes \mathds{1}
+ u_l^{(2)}(t)\, \mathds{1} \otimes \sigma_l
\right),
\label{eq:H_kk}
\end{align}
where $k,l \in \{x,y,z\}$. For $J_k\neq 0$, the fixed interaction $J_k(\sigma_k\otimes\sigma_k)$, combined with independent local controls, can generate arbitrary single-qubit rotations and is sufficient for universal two-qubit control, including the synthesis of the CX gate~\cite{Dodd2002}.

\subsection{Objective gates: Hadamard and Controlled-NOT}

In this section, we briefly introduce the single-qubit Hadamard and the two-qubit CX gates. These two unitary operations are of great value in quantum computing, given that Hadamard allows superposition of quantum states of a qubit, while CX leads to entanglement in multipartite setups. Hence, both are essential to universal quantum computation based on elementary gates.

The Hadamard gate is a single-qubit operation that creates an equal superposition of computational basis states; thus, it is sensitive to phase damping noise, which destroys quantum superposition. In the computational basis $\{\ket{0}, \ket{1}\}$, Hadamard is defined by the unitary matrix

\begin{equation}
H = \frac{1}{\sqrt{2}}
\begin{pmatrix}
1 & 1 \\
1 & -1
\end{pmatrix},
\end{equation}

which transforms the computational basis states as
\begin{equation}
H\ket{0} = \frac{\ket{0} + \ket{1}}{\sqrt{2}}, \qquad
H\ket{1} = \frac{\ket{0} - \ket{1}}{\sqrt{2}}.
\end{equation}
Physically, the Hadamard gate can be realized by a suitable sequence of rotations around the
$x$ and $z$ axes of the Bloch sphere, or equivalently by a single properly shaped control pulse
generated through the Hamiltonian \cref{eq:SQH_gen}.

The CX gate is a two-qubit entangling operation defined by
\begin{equation}
\text{CX}_{c,t} =
\begin{pmatrix}
1 & 0 & 0 & 0 \\
0 & 1 & 0 & 0 \\
0 & 0 & 0 & 1 \\
0 & 0 & 1 & 0
\end{pmatrix},
\label{eq:CX-gate}
\end{equation}
where the first qubit is the control ($c$) and the second is the target ($t$). Its action can be written as
\begin{equation}
|c,t\rangle \mapsto |c, t \oplus c\rangle, \label{eq:cond_rule}
\end{equation}
where $c,t \in \{0,1\}$ stand for control and target, respectively. The CX gate cannot be decomposed into independent single-qubit operations and therefore requires qubit-qubit couplings, as in \cref{eq:H_kk}. Generating a high-fidelity CX gate depends on a quantum platform's ability to implement a controlled entangling operation. 

\begin{figure*}[tbp]
    \centering
    \includegraphics[width=1\linewidth]{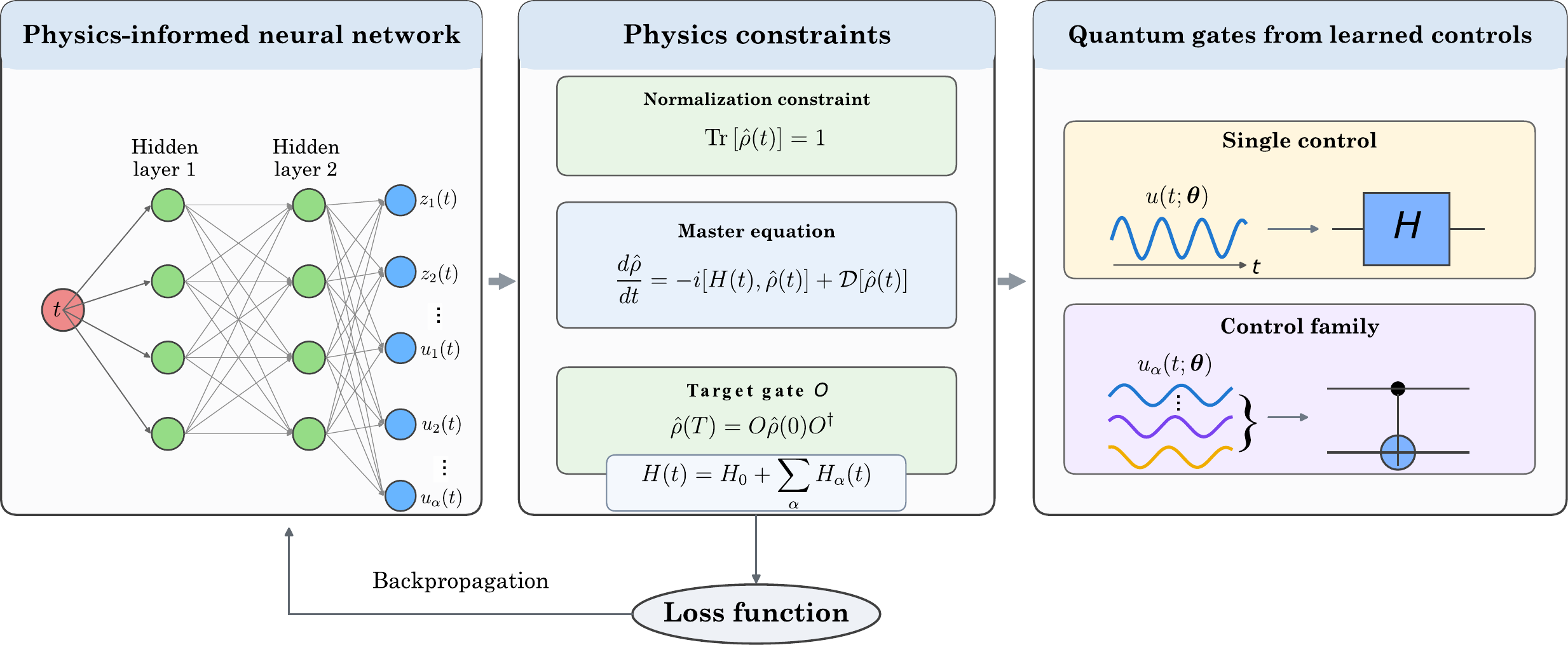}
    \caption{Overview of the PINN framework
for designing one- and two-qubit quantum gates via inverse optimal
control. The network takes time $t$ as input and produces
$\boldsymbol{N}_z(t)$ and $\boldsymbol{N}_u(t)$, which parametrize the
density-matrix components $\boldsymbol{z}(t)$ and control fields
$\boldsymbol{u}(t)$ through the ansatz functions $f(t)$ and $\Theta(t)$,
respectively, incorporating the initial-state and pulse boundary
conditions. The control fields enter the time-dependent Hamiltonian
$H(t)=H_0+\sum_{\alpha}u_{\alpha}(t)H_{\alpha}$.
Training minimizes a loss function combining the unitary
density-matrix dynamical residual, the unit-trace constraint, a target-gate
term, and network-parameter regularization. The target-gate term
penalizes deviations from
$\hat{\rho}^{(s)}(T)=O\hat{\rho}^{(s)}(0)O^{\dagger}$
for the initial states indexed by $s$ in the training set,
using the same control fields for all states.
The feedback loop represents the update of the network parameters
through backpropagation. The right panel illustrates a single
control field for the Hadamard gate and a family of control fields
for the controlled-NOT (CX) gate.
}
    \label{fig:PINN}
\end{figure*}

\section{Physics-informed neural networks for quantum gates}
\label{sec:PINN}

\subsection{General operating principle}

PINNs provide a unified framework for solving forward and
inverse problems governed by differential equations by embedding the physical laws directly into the training process of neural networks. Unlike purely data-driven approaches, PINNs constrain learning by enforcing the governing equations as soft constraints in the loss function. This promotes consistency between the learned solutions and the underlying physics, even in regimes where data are sparse or noisy.

In the context of quantum control, a PINN is designed to approximate both the quantum state (or unitary propagator) and the control functions as outputs of a neural network parameterized
by a set of weights and biases. The unitary dynamics can be incorporated into a loss function through a residual term that penalizes deviations between a desired state and one simulated through the Schr\"odinger equation~\eqref{eq:unitary_u}. Moreover, we can construct an ansatz that enforces the initial and pulse boundary conditions~\cite{Norambuena2024}, while additional loss terms penalize target-gate errors, trace deviations, and large network weights. We propose the following loss function for quantum gates
\begin{equation}
\mathcal{L} = \mathcal{L}_{\text{dyn}} + \mathcal{L}_{\text{gate}} + \mathcal{L}_{\text{trace}} +
\mathcal{L}_{\text{reg}}, \label{eq:loss}
\end{equation}
where $\mathcal{L}_{\text{dyn}}$ enforces the unitary quantum dynamics of the system, $\mathcal{L}_{\text{gate}}$ enforces convergence toward the desired unitary operation
(e.g., Hadamard or CX), $\mathcal{L}_{\text{trace}}$ penalizes deviations from unit trace, and $\mathcal{L}_{\text{reg}}$ regularizes the network weights.

The scheme of our PINN model for solving the inverse problem in quantum control for quantum gates follows the structure outlined in Ref.~\cite{Norambuena2024}. We use the notation $\boldsymbol{z}(t)=\operatorname{vec}_{\mathbb{R}}(\hat{\rho}(t)) \in \mathbb{R}^{d^2}$ which defines a real-valued vector whose components are given by $\boldsymbol{z}(t) =
\big(
\{\rho_{ii}(t)\}_{i=0}^{d-1},\;
\{\mathrm{\Re}(\rho_{ij}(t)),\, \mathrm{\Im}(\rho_{ij}(t))\}_{0\leq i<j \leq d-1}
\big)^{T}$, where $\rho_{ij}(t)=\bra{i}\hat{\rho}(t)\ket{j}$ are the elements of the density matrix which satisfies the Hermiticity condition $\rho_{ji}(t) = \rho_{ij}^*(t)$, so that only the upper triangular elements are treated as independent variables. In the case of a qubit, $d=2$ so $0 \leq i<j\leq 1$, and the component vector reduces to $\boldsymbol{z}(t) =
\big(
\rho_{00}(t), \rho_{11}(t),\;
\mathrm{\Re}(\rho_{01}(t)),\, \mathrm{\Im}(\rho_{01}(t))
\big)^{T}$. We define the control vector as $\boldsymbol{u}(t)=(u_1(t),\cdots, u_\alpha(t))^T \in \mathbb{R}^\alpha$, where $\alpha \geq 1$ denotes the number of control functions. In the context of our problem, a quantum gate corresponds to a unitary transformation that must act consistently on any input state \cite{nielsen2010quantum}. Consequently, we consider multiple target states simultaneously, defined as $\boldsymbol{z}^{(s)}_{\mathrm{d}} \in \mathbb{R}^{d^2}$, where $s=1,\ldots,N_{\mathrm{tr}}$ indexes the training inputs and $\boldsymbol{z}_{\mathrm{d}}^{(s)}$ represents the desired state associated with $\hat{\rho}^{(s)}(0)$ through $O\hat{\rho}^{(s)}(0)O^\dagger$. The same control fields are shared by all $N_{\mathrm{tr}}$ state trajectories.

The proposed PINN architecture for quantum gates is shown in \cref{fig:PINN}, which is supported by the universal approximation theorem, which guarantees that feed-forward neural networks with sufficient capacity (appropriate network parameters and non-linear activation functions) can approximate any continuous function in compact domains with arbitrary precision~\cite{hornik1991approximation}. This provides an approximation basis for representing time-dependent trajectories and controls; consistency with the dynamics must still be assessed through training and independent propagation. In the structure illustrated in \cref{fig:PINN}, our PINN takes as input a set of temporal collocation points $\boldsymbol{t} = (t_1,\cdots,t_{N_c})^T \in \mathbb{R}^{N_c}$, is fully connected with $(l-1)$ hidden layers of $m$ neurons per layer, and has an output layer $l$. Here, $N_c$ denotes the number of collocation points used to discretize the temporal domain during training. Additionally, each perceptron in the hidden layers uses a sinusoidal activation function $\sigma=\sin(\cdot)$, introducing nonlinearity and enabling the representation of complex temporal features in both the system dynamics and control functions. We further analyze the effects of eight activation functions and three optimizers in Appendix~\ref{app:CX_PINN_robustness}. These comparisons show that the best choices depend on the target gate and the evaluation metric: lower training loss does not necessarily mean higher gate fidelity. We therefore assess gate performance by independently propagating the learned control fields.

The network outputs $\boldsymbol{N}_z^{(s)}(t)$ and $\boldsymbol{N}_u(t)$ parametrize the state trajectories and the shared controls through
\begin{align}
\boldsymbol{z}^{(s)}(t)&=\boldsymbol{z}^{(s)}(0)+f(t)\boldsymbol{N}_z^{(s)}(t),\nonumber\\
\boldsymbol{u}(t)&=\Theta(t)\boldsymbol{N}_u(t),
\label{eq:pinn_ansatz}
\end{align}
where $f(t)=1-e^{-t}$ and $\Theta(t)=(1-e^{-t})^n(1-e^{t-T})^n$, using the numerical time units of the simulations. The controls satisfy $\boldsymbol{u}(0)=\boldsymbol{u}(T)=0$. The integer $n\geq1$ sets the endpoint smoothness: derivatives of $\Theta$ of orders $k=0,\ldots,n-1$ vanish at both endpoints. We use $n=2$, so both the control amplitudes and their first derivatives vanish at $t=0$ and $t=T$. Here, $T$ is the gate duration; fixed drift terms remain present when the excitation controls are switched off.

In quantum platforms, control operations use $\pi$ and $\pi/2$ resonant pulses, which correspond to time-dependent electromagnetic fields, as demonstrated in solid-state qubit systems such as color centers in diamond~\cite{PhysRevLett.119.260501}. In neutral-atom architectures, these controls are implemented via laser pulses with a time-modulated Rabi frequency $\Omega(t)$, applied over a finite time interval \cite{evered2023high}. As a result, the control fields are intrinsically time-localized and are switched on and off during the operation.

We now describe each term in the loss function in \cref{eq:loss}. The physics loss term $\mathcal{L}_{dyn}$ is defined as:
\begin{equation}
    \mathcal{L}_{dyn}=\sum_{k=1}^{N_{\mathrm{tr}}}\frac{1}{N_c}\sum_{i=1}^{N_c}\vert\vert \dot{\boldsymbol{z}}^{(k)}(t_i) - \mathscr{L}(\boldsymbol{z}^{(k)}(t_i))\vert\vert^{2}_{2},
\end{equation}
where $\Vert\cdot\Vert_2^2$ is the squared Euclidean norm and $\mathscr{L}$ is the real-coordinate representation of $-i[H(t),\hat{\rho}(t)]$. All reported controls are trained with this unitary residual. For the PINN, this term represents the residual of a set of coupled differential equations parametrized by $\boldsymbol{z}(t)$. The target-gate term $\mathcal{L}_{gate}$ is written as:
\begin{equation}
    \mathcal{L}_{gate}= \eta\sum_{k=1}^{N_{\mathrm{tr}}} \left\vert\left\vert \boldsymbol{z}^{(k)}(T) - \boldsymbol{z}^{(k)}_{\mathrm{d}}\right\vert \right\vert^{2}_{2},
\end{equation}
where the weight $10^{-3}\leq \eta\leq 1$ controls the contribution of $\mathcal{L}_{gate}$ with respect to $\mathcal{L}_{dyn}$, ensuring that the learned dynamics are consistent with the desired unitary transformation across simultaneous target state configurations. This term is closely related to the target-gate fidelity we test outside the training setup. We also consider a constraint term to enforce the unit-trace property of the density operator ($\mathrm{Tr}\left[\hat{\rho}(t)\right] = 1$), which in terms of the vector $\boldsymbol{z}(t)$ reduces to
\begin{equation}
    \mathcal{L}_{trace}=\sum_{k=1}^{N_{\mathrm{tr}}}\frac{1}{N_c}\sum_{i=1}^{N_c}\left\vert 1 - \sum_{j=0}^{d-1} {\rho}^{(k)}_{jj}(t_i) \right\vert^2.
\end{equation}

Finally, the regularization term $\mathcal{L}_{\mathrm{reg}}=\chi\Vert\boldsymbol{w}\Vert_2^2$ penalizes the network weights $\boldsymbol{w}$, with strength $\chi$, to mitigate overfitting \cite{Zhao2025Jul}. This penalty can influence the learned control amplitudes indirectly, but does not impose a hard amplitude or bandwidth bound. The pulse boundary smoothness follows from the ansatz in Eq.~\eqref{eq:pinn_ansatz}.

Minimizing $\mathcal{L}$ provides state trajectories and smooth control pulses for the specified target gate and Hamiltonian. The quality of the resulting gate is assessed independently by propagating the learned controls and evaluating fidelities on input states outside the training set.

\section{Results}
\label{sec:result}

In this section, we evaluate the performance of the PINN-derived control fields for both single- and two-qubit quantum gates. First, we assess performance by analyzing the statistical properties of quantum fidelity (mean and standard deviation) using Monte Carlo sampling over random initial states not considered during training. We also calculate the average gate fidelity using the Nielsen formula for unitary evolution. In the weak-dissipation regime, we explicitly show that the fidelity deteriorates linearly with the damping rates for both amplitude and phase damping channels. Finally, we separately analyze the effects of the bilinear couplings $\sigma_i \otimes \sigma_i$ for the two-qubit gates.

\subsection{PINN-predicted control for the single-qubit Hadamard operation}

The first task for our PINN to reproduce is the single-qubit Hadamard gate. To do so, we consider the Hamiltonian \cref{eq:SQH} and a set of four initial states given by:

\begin{equation}
    \left\{ \ket{\psi_j(0)} \right\}_{j=1}^4=\{\ket{0}, \ket{1}, \ket{+},\ket{-}\},
    \label{eq:init-Hadamard}
\end{equation}

that corresponds to a set of two mutually unbiased bases: the computational basis $\{\ket{k^{(1)}}\}=\{\ket{0}, \ket{1}\}$, and the Hadamard basis $\{\ket{k^{(2)}}\}=\{\ket{+}, \ket{-}\}$, respectively. For a $d$-dimensional Hilbert space, two orthonormal bases $\{| k^{(1)}_i \rangle\}_{i=1}^d$ and $\{| k^{(2)}_i \rangle\}_{i=1}^d$ are mutually unbiased when \cite{goerz2014optimal, reich2013minimum}:
\begin{equation}
    \left|\braket{k^{(1)}_i}{k^{(2)}_j}\right|^2
    =\frac{1}{d},
    \qquad \forall\, i,j. \label{eq:unbiased}
\end{equation}

\begin{figure}[tbp]
    \centering
    \includegraphics[width=\linewidth]{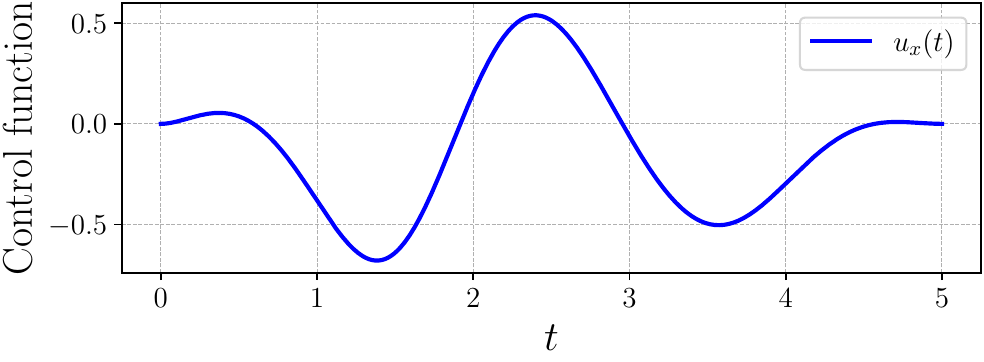}\\

    \includegraphics[width=0.98\linewidth]{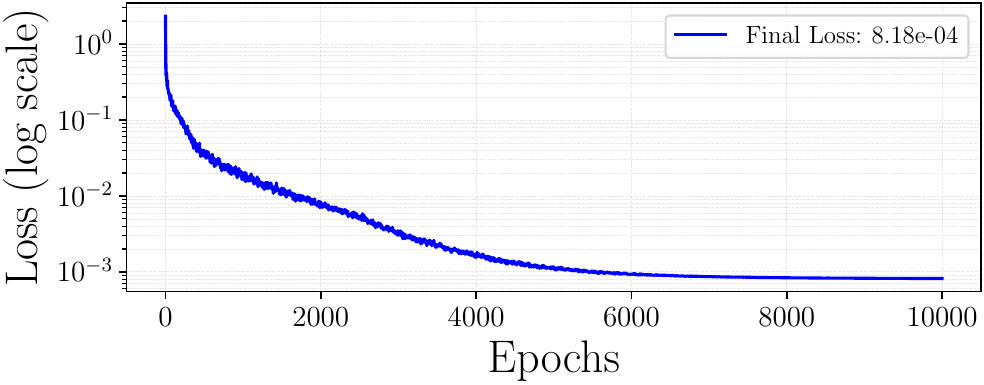}

    \caption{Hadamard-gate control function predicted by PINN. (Top) Shape of the control function satisfying initial and final time conditions described in Sec. \ref{sec:PINN}. PINN architecture: $6$ hidden layers of $300$ neurons each, optimized with $10^4$ training epochs with a learning rate of $10^{-3}$, $\eta = 10^{-1}$ and $\chi = 10^{-5}$. (Bottom) Total loss function \cref{eq:loss} in terms of epochs.}
    \label{fig:CF_single_H}
\end{figure}

Figure~\ref{fig:CF_single_H} shows the learned control function predicted by our PINN scheme, considering a final evolution time of $T=5$.
It can be observed that the control function exhibits a smooth, well-behaved, continuous, and differentiable time profile. This smoothness reflects the PINN's ability to approximate physically stable solutions that simultaneously satisfy the system dynamics and the boundary conditions imposed by the target operation. In particular, the smoothness of the pulse indicates that the Hadamard gate can be implemented using stable control trajectories in state space, avoiding the need for rapid or high-frequency modulations. Details of the training convergence and numerical robustness are presented in \cref{app:H_PINN_robustness}.

A lower training loss does not always imply a higher gate fidelity. This means that lower values of the total loss function might end up in a worse fidelity for the target gate. Hence, our figure of merit to evaluate the performance of the control functions predicted by our PINN is the mean fidelity over states outside the training set. In this work, we use the average state fidelity over an ensemble of random pure states, defined as:

\begin{equation}
    F_m=
    \frac{1}{N_s}
    \sum_{i=1}^{N_s}
    \bra{\phi_i}\hat{\rho}_i(T)\ket{\phi_i},
    \label{eq:Fm}
\end{equation}
where $\hat{\rho}_i(T)$ is the independently propagated final density operator for the initial pure state $\ket{\psi_i(0)}$ under the learned control. For unitary evolution, $\hat{\rho}_i(T)=\ketbra{\psi_i(T)}{\psi_i(T)}$, and each summand reduces to $|\braket{\phi_i}{\psi_i(T)}|^2$. On the other hand, $\ket{\phi_i}=O\ket{\psi_i(0)}$ is the ideal state obtained by the implementation of the target gate $O$ on the same initial state. The initial pure states are generated randomly according to the Haar measure in Hilbert space using the QuTiP library \cite{johansson2012qutip, PhysRevA.71.032313}, and $N_s$ denotes the number of random initial states.

In the limit of a sufficiently large number of sampled states, this Monte Carlo estimate approaches the average fidelity over the complete set of pure input states
\begin{equation}
    F_{\mathrm{avg}}(\mathcal{E}, O)  =
    \int d\psi\, \bra{\psi} O^\dagger \mathcal{E}\left(\psi\right)O\ket{\psi},
    \label{eq:Favg_integral}
\end{equation}
where $\psi=\ketbra{\psi}{\psi}$, $\mathcal{E}(\cdot)$ is a completely positive, trace-preserving quantum channel, $O$ is the unitary operator corresponding to the ideal
gate, and the integral is taken over the uniform (Haar) measure $d\psi$ on the pure state space. For a quantum operation acting on a Hilbert space of dimension $d$, the average gate fidelity defined in Eq.~\eqref{eq:Favg_integral} can be written as \cite{nielsen2002simple}:
\begin{equation}
    F_{\mathrm{avg}}(\mathcal{E},O) =
    \frac{
    \displaystyle{\sum_{j} \tr\left( O B_j^\dagger O^\dagger \mathcal{E}(B_j) \right) +d^2}}
    {d^2(d+1)},
    \label{eq:Nielsen}
\end{equation}

where $\{B_j\}_{j=1}^{d^2}$ is a complete basis of unitary operators satisfying $\tr[B_j^{\dagger}B_k]=d\delta_{jk}$. Examples of such operator bases are given in \cite{knill1996non}. For unitary dynamics acting on the set of operators $B_j$, we have
$\mathcal{E}(B_j)=U(T)B_jU^{\dagger}(T)$. Here, $U(T)$ denotes the unitary propagator generated by the controlled dynamics at the final time $T$.

\begin{figure}[tbp]
    \centering
    \vspace{0cm}

    \begin{overpic}[width=\columnwidth]{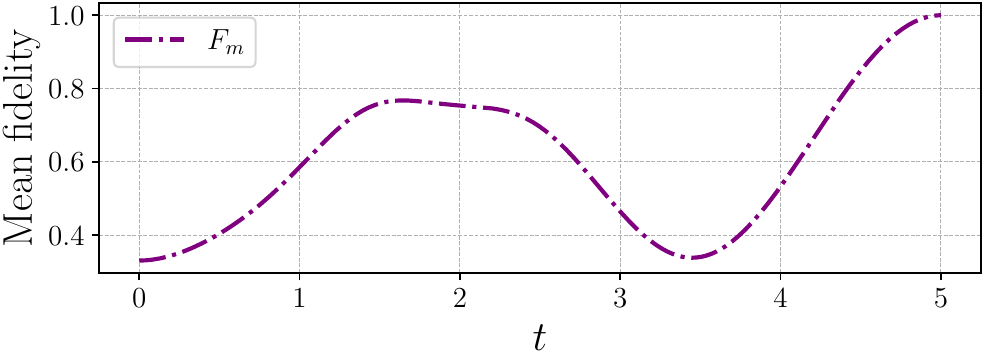}
        \put(15,15){\small\text{(a)}}
    \end{overpic}

    \vspace{0cm}

    \begin{overpic}[width=\columnwidth]{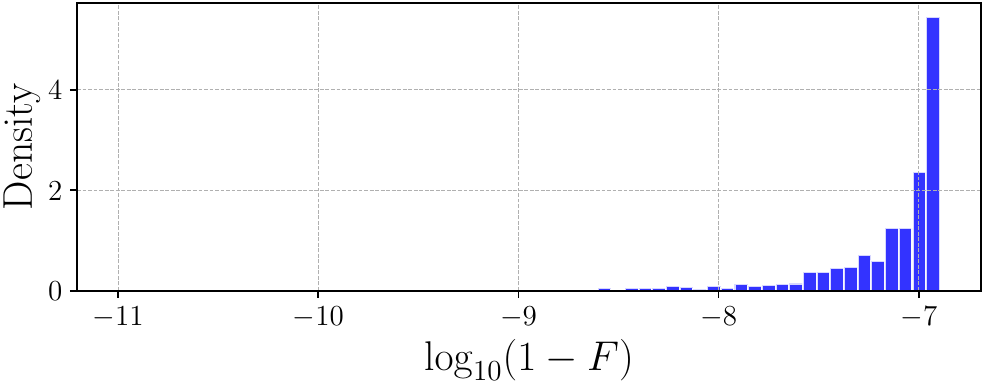}
        \put(14,12){\small\text{(b)}}
    \end{overpic}

    \vspace{0cm}

    \begin{overpic}[width=\columnwidth]{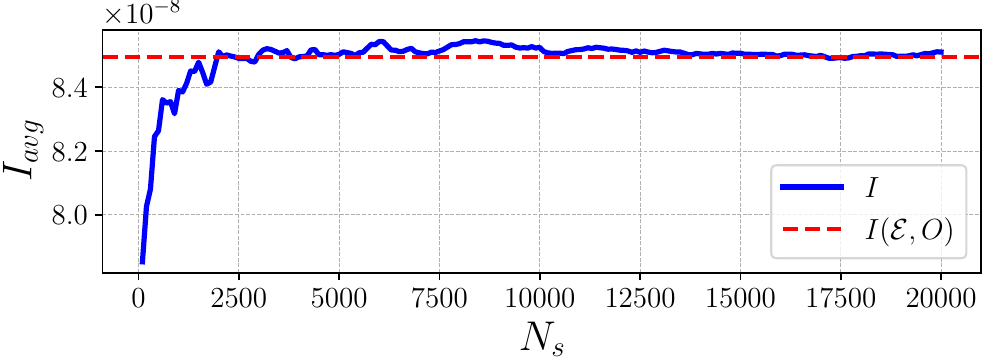}
        \put(15,13){\small\text{(c)}}
    \end{overpic}

    \caption{PINN-assisted Hadamard fidelity. (a) Average fidelity as a function of time for the Hadamard gate, evaluated by \cref{eq:Fm}. (b) Monte Carlo distribution of the gate infidelity through the exponent $\log_{10}(1-F)$ and evaluated at the final evolution time $T$; average infidelity $I_{\mathrm{avg}} = 8.59\times10^{-8}$, standard deviation $I_{\mathrm{std}}=3.72\times10^{-8}$, minimum infidelity $I_{\mathrm{min}}=1\times10^{-11}$, and maximum infidelity $I_{\mathrm{max}}=1.27\times10^{-7}$. (c) Behavior of the average infidelity $I=1-F_m$ as $N_s$ increases. The red-dashed line is the average gate infidelity $I(\mathcal{E},O)=1-F_{\mathrm{avg}}(\mathcal{E},O)=8.49 \times10^{-8}$.}
     \label{fig:Fm_single_H}
\end{figure}

Figure~\ref{fig:Fm_single_H} shows the fidelity behavior associated with the dynamics given by the PINN-predicted control function. Panel (a) shows the time evolution of the average fidelity for the Hadamard gate, evaluated using $N_s=10^3$ random initial states. Throughout the evolution, the average fidelity follows a smooth temporal behavior and progressively approaches the target operation, reaching its maximum value at the final evolution time $T$, with $F_m\approx 99.9999914\%$. To further characterize the quality of the obtained gate, panel (b) shows the Monte Carlo distribution of the gate infidelity, represented through the exponent $\log_{10}(1-F)$, evaluated at the final time. The concentration of the samples around low infidelity values demonstrates that the optimized control provides a consistent implementation of the Hadamard gate over a broad set of sampled pure states distributed by the Haar measure. This trend is reflected in the statistical measures of the distribution, represented by an average infidelity $I_{\mathrm{avg}} = 8.59\times10^{-8}$, a standard deviation $I_{\mathrm{std}} =3.72\times10^{-8}$, and a range of ($1\times10^{-11}$, $1.27\times10^{-7}$). Although the standard deviation is comparable to the mean, the absolute infidelities remain small across the sampled input states. Finally, panel (c) shows how the average infidelity (taken from \cref{eq:Fm}) converges to the \textit{reference} average infidelity extracted from \cref{eq:Nielsen}, as the number of sampled pure states $N_s$ increases.

To assess sensitivity to environmental noise, we reuse the controls trained under unitary dynamics and solve the Lindblad master equation
\begin{align}
\dot{\hat{\rho}}&=-i[H(t),\hat{\rho}]+\sum_k\gamma_k\mathcal{D}[L_k]\hat{\rho},\nonumber\\
\mathcal{D}[L]\hat{\rho}&=L\hat{\rho}L^\dagger-\tfrac12\{L^\dagger L,\hat{\rho}\}.
\label{eq:lindblad}
\end{align}
Here, $\gamma_k$ is the coefficient multiplying the normalized jump operator's dissipator; the operator normalization fixes the relation to physical relaxation or dephasing times. The unitary controls are not retrained in the presence of noise. For $\gamma_kT\ll1$, the average fidelity of an otherwise ideal gate admits the first-order expansion~\cite{abad2022universal,abad2025impact}
\begin{equation}
    F(\mathcal{E}, O)
    = 1 + \sum_k \gamma_k \int_0^T\delta F\left(L_k(t)\right)\,dt + \mathcal{O}\left((\gamma T)^2\right),
    \label{eq:first_order_fidelity}
\end{equation}
where $L_k$ denotes the Lindblad jump operators associated with the $k$-th dissipative process, and $\delta F(L_k(t))$ quantifies its first-order contribution to the fidelity reduction. For quantum operations confined to the computational subspace, $\delta F(L_k(t))$ admits a gate-independent analytical expression that depends on the system dimension, the gate duration, and the corresponding dissipative channel \cite{abad2022universal, abad2025impact}. In particular, for an $N$-qubit system with local Lindblad operators acting on individual qubits, the first-order contribution can be written explicitly for specific dissipative channels. For a single local pure-dephasing channel with Lindblad operator $L_z=\sigma_z$ and rate $\gamma_z$, the average gate fidelity takes the form
\begin{equation}
    F_z(\mathcal{E}, O)= 1+\gamma_z\delta F(L_z\otimes\mathds{1}_{N-1})T, \label{eq:Fdissipative_lz}
\end{equation}
whereas, for amplitude damping with Lindblad operator $L_-=\sigma_-$ and rate $\gamma_-$ (spontaneous emission), the corresponding expression is
\begin{equation}
    F_-(\mathcal{E}, O)= 1+\gamma_-\delta F(L_-\otimes\mathds{1}_{N-1})T, \label{eq:Fdissipative_l-}
\end{equation}
where $\mathds{1}_{N-1}$ is the identity on the other $N-1$ qubits and $\delta F(L_z\otimes\mathds{1}_{N-1})=2\delta F(L_-\otimes\mathds{1}_{N-1})=-d/(d+1)$. We remark that the spontaneous emission process is valid at low temperatures, $k_B T_{\mathrm{bath}} \ll \hbar \omega_0$, where $T_{\mathrm{bath}}$ is the bath temperature and $\hbar \omega_0$ is the qubit energy scale. For higher bath temperatures, one must include the absorption mechanism, \textit{i.e.}, $L_+=\sigma_+$ with rate $\gamma_+$. In the case of a qubit, \cref{eq:Fdissipative_lz,eq:Fdissipative_l-} simplify to $F_z(\mathcal{E}, O)=1-(2/3)\gamma_zT$ and $F_-(\mathcal{E}, O)=1-(1/3)\gamma_-T$, respectively, showing the linear reduction in average gate fidelity with the decay rate.  For operations that evolve outside the computational subspace, this
simplification no longer applies; in this case, $\delta F(L_k(t))$ depends explicitly on the time-evolved jump operator $L_k(t)=U^\dagger(t)L_kU(t)$ and is evaluated using the generalized
expression derived in \cite{abad2025impact}.

\begin{figure}[tbp]
    \centering
        \includegraphics[width=\linewidth]{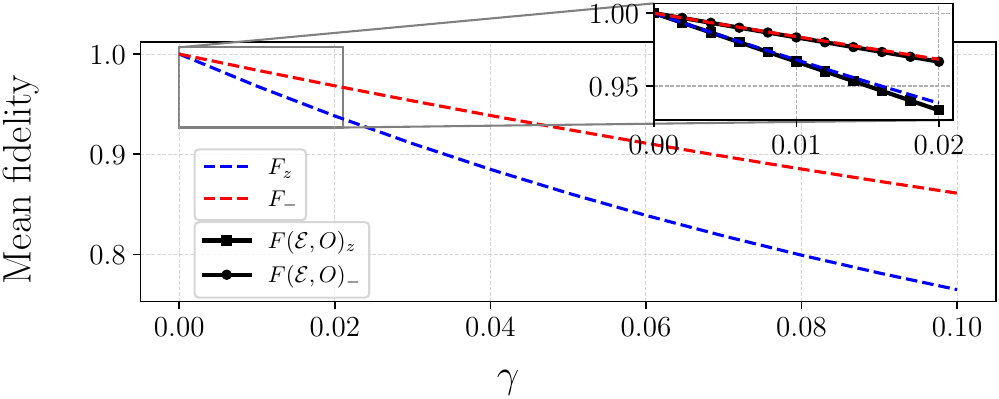}

    \caption{Average fidelity under noisy channels for Hadamard. \cref{eq:Fm} computed for the Hadamard gate considering pure dephasing (blue curve) and amplitude damping (red curve), as a function of the decay rate $\gamma/\Delta \in [0,10^{-1}]$. The inset shows a zoom of the weak-dissipation regime ($\gamma T \ll1$), where $F_z(\mathcal{E},O)$ and $F_-(\mathcal{E},O)$ correspond to the average gate fidelities according to \cref{eq:Fdissipative_lz,eq:Fdissipative_l-} respectively.}
    \label{fig:Infidelity_PD_AD_Hadamard}
\end{figure}

In Figure~\ref{fig:Infidelity_PD_AD_Hadamard}, we show the behavior of the average fidelity of the PINN-assisted Hadamard implementation under pure dephasing (blue curve) and amplitude damping (red curve) noisy channels. Each curve is obtained by extracting the control function from the PINN trained in the unitary case ($\gamma=0$), and then solving the Markovian quantum dynamics for a fixed value of $\gamma $ with the control function in the Hamiltonian. Finally, the average fidelity at the final time is computed over $N_s = 10^3$ random initial states for each value of $\gamma$. As expected, the average fidelity decreases monotonically with the decay rate $\gamma$ for both noisy channels, showing the progressive degradation induced by the interaction with the environment. For the jump-operator normalizations in Eq.~\eqref{eq:lindblad} and equal numerical rates, pure dephasing produces a larger fidelity reduction than amplitude damping, consistent with their respective first-order coefficients. The inset shows the behavior of our results in the weaker regime ($\gamma T \ll1$), where symbols illustrate the corresponding first-order approximation given by \cref{eq:Fdissipative_lz,eq:Fdissipative_l-} for the average fidelity $F_z$ and $F_-$, respectively. The match among symbols and curves corroborates the linear dependence of the fidelity with respect to the decay rate.

\subsection{PINN-predicted controls for the two-qubit CX gate}
\label{sec:PINN_tqCX}

We now extend the analysis to two-qubit systems and investigate the performance of the control functions predicted by the PINN to implement the CX gate, considering longitudinal and transverse couplings between the qubits. The PINN training hyperparameters are held constant across all qubit-qubit couplings, so that all coupling configurations are compared using the same network architecture and training protocol. In particular, our PINN architecture consists of $5$ hidden layers and $300$ neurons per layer, optimized over $2.5\times10^4$ epochs, with a learning rate of $1\times10^{-3}$, a regularization hyperparameter of $1\times10^{-4}$, and a weight of $1\times10^{-1}$ for each target.

\begin{figure}[tbp]
    \centering
    \vspace{0.1cm}

    \begin{overpic}[width=\columnwidth]{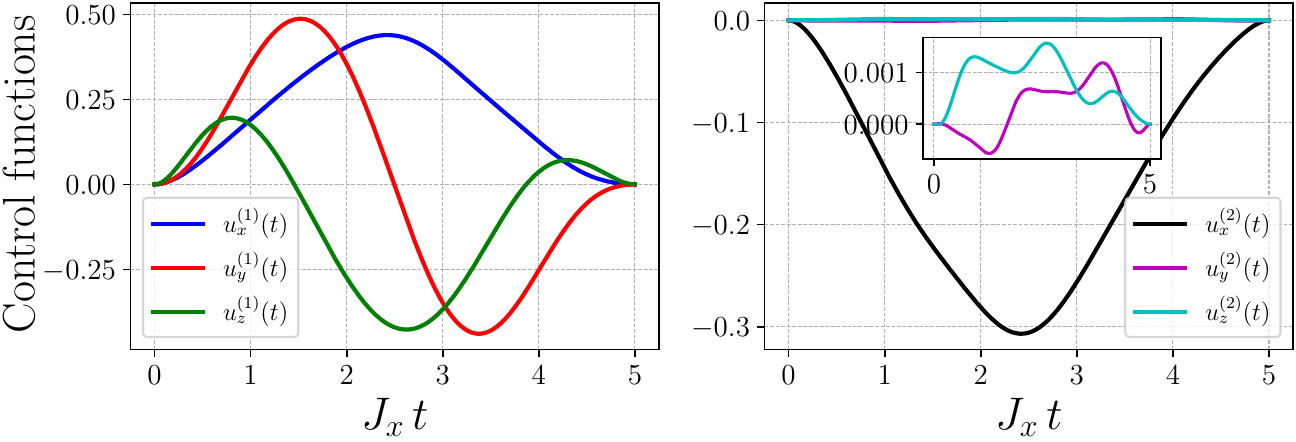}
        \put(37,36){\small\text{(a) $\sigma_x\otimes\sigma_x$ coupling.}}
    \end{overpic}

    \vspace{0.40cm}

    \begin{overpic}[width=\columnwidth]{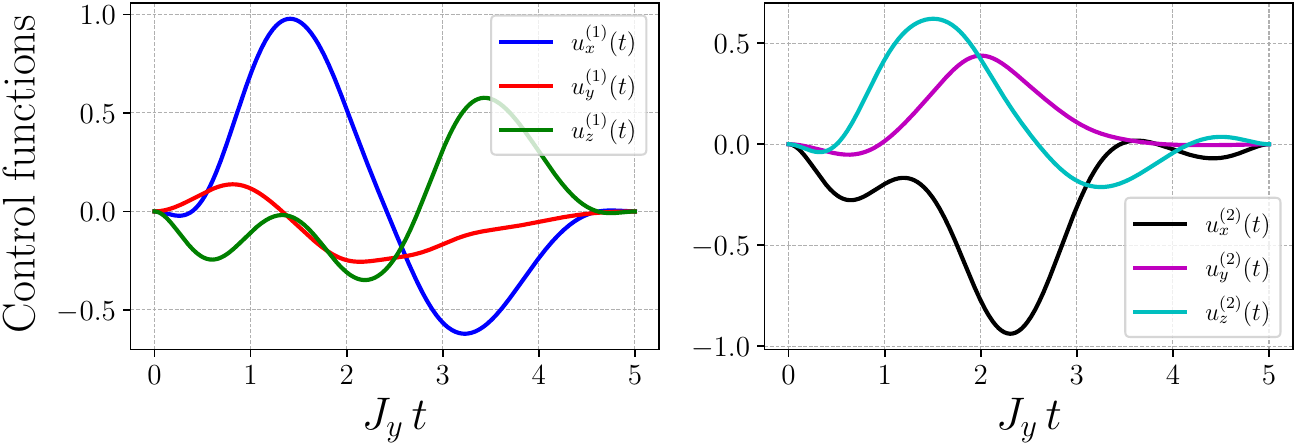}
        \put(37,36){\small\text{(b) $\sigma_y\otimes\sigma_y$ coupling.}}
    \end{overpic}

    \vspace{0.40cm}

    \begin{overpic}[width=\columnwidth]{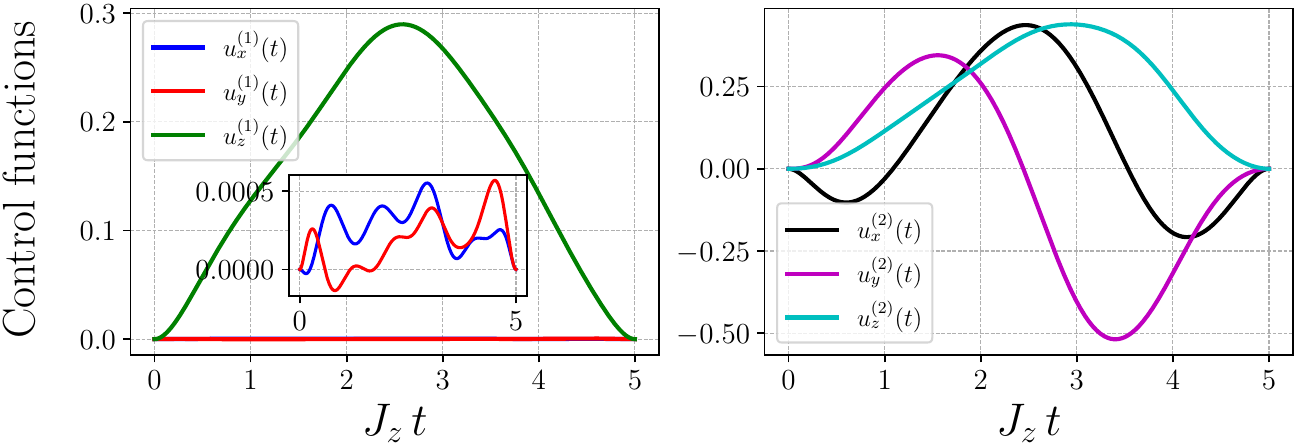}
        \put(37,36){\small\text{(c) $\sigma_z\otimes\sigma_z$ coupling.}}
    \end{overpic}

    \caption{Local control functions for the CX gate. (a) $\sigma_x\otimes\sigma_x$, (b) $\sigma_y\otimes\sigma_y$ and (c) $\sigma_z\otimes\sigma_z$. Left panels show functions on the control qubit (qubit 1) and right panels on the target qubit (qubit 2). Insets show a zoom of the control functions that are at least two orders of magnitude smaller. The set of initial states is $\mathcal{S}_3$.}
    \label{fig:CF_CNOT_S3}
\end{figure}

The six time-dependent local control functions predicted by the PINN for the CX gate under different qubit--qubit couplings are shown in \cref{fig:CF_CNOT_S3} as follows: couplings (a) $\sigma_x\otimes\sigma_x$, (b) $\sigma_y\otimes\sigma_y$ and (c) $\sigma_z\otimes\sigma_z$ are considered separately to analyze the effects on the predicted functions. The three control functions acting on the control qubit  are shown in the left panels while those acting on the target qubit are plotted in the right panels.

The set of initial states used to train the PINN for the CX gate implementation follows the same structure of unbiased bases described in the Hadamard case. For this two-qubit scenario, we consider three distinct 8-state sets as follows: $\mathcal{S}_1 = \{|k^{(1)}\rangle \cup |k^{(3)}\rangle\}$, $\mathcal{S}_2 = \{|k^{(1)}\rangle \cup |k^{(2)}\rangle\}$, and $\mathcal{S}_3 = \{|k^{(2)}\rangle \cup |k^{(4)}\rangle\}$, where $\ket{k^{(1)}}, \ket{k^{(2)}}, \ket{k^{(3)}}, \ket{k^{(4)}}$ are given by:

\begin{align}
    &|k^{(1)}\rangle=\{\ket{00},\ket{01},\ket{10},\ket{11}\}, \nonumber\\
    &|k^{(2)}\rangle=\{\ket{++},\ket{+-},\ket{-+},\ket{--}\},\nonumber\\
    &|k^{(3)}\rangle= \{\ket{\chi_{00}},\ket{\chi_{01}},\ket{\chi_{10}},\ket{\chi_{11}}\} \nonumber,\\
    &|k^{(4)}\rangle=\{\ket{\Phi_{+i}},\ket{\Phi_{-i}}, \ket{\Psi_{+i}}, \ket{\Psi_{-i}}\}. \nonumber
\end{align}

$\ket{k^{(1)}}$  and $\ket{k^{(2)}}$ stand for the computational and Hadamard bases, respectively. $\ket{k^{(3)}} = \{\ket{\chi_{ij}}=CZ|k^{(2)}_{ij}\rangle\}$ is obtained after applying the controlled-phase (CZ) gate on the Hadamard basis, and $\ket{k^{(4)}}$ is composed of the entangled states $\ket{\Phi_{\pm i}}=\tfrac{1}{\sqrt{2}}(\ket{00}\pm i\ket{11})$ and $\ket{\Psi_{\pm i}}=\tfrac{1}{\sqrt{2}}(\ket{01}\pm i\ket{10})$. \cref{fig:CF_CNOT_S3} shows the results for $\mathcal{S}_3$.

Our results show that a CX gate with high fidelity might be simulated with every considered coupling ($\sigma_x\otimes\sigma_x \, \rightarrow \, F_{\mathrm{avg}}(\mathcal{E},O)\approx99.99997\%$, $\sigma_y\otimes\sigma_y \, \rightarrow \, 99.99951\%$ and $\sigma_z\otimes\sigma_z \, \rightarrow \, 99.99988\%$). However, the local control differs in each case. In particular, for the coupling along $\sigma_x$, the learned solution has three appreciable controls on the control qubit and one dominant control on the target qubit. The function $u_x^{(2)}$ (black curve in the right panel of \cref{fig:CF_CNOT_S3}(a)) is about two orders of magnitude greater than the functions along $y$ and $z$. A similar behavior is observed for the coupling along $\sigma_z$ but now requiring a single control ($u_z^{(1)}$) on the control qubit. This can be explained from the block structure of the CX gate  \cref{eq:CX-gate}, which is diagonal for the control qubit in the computational basis $\{\ket{0}, \ket{1}\}$ while diagonal in the Hadamard basis $\{\ket{+}, \ket{-}\}$ for the target qubit. Hence, the \textit{opposite} behavior exhibited by the control functions for these two couplings reflects the fact that a direction is favored by the system to organize the local control according to the considered coupling. This is also in agreement with the control functions predicted for the $\sigma_y\otimes\sigma_y$ coupling. In this case, neither the computational basis nor the Hadamard one is favored by the coupling, and the learned solution has six appreciable components. This observation does not establish a minimum number of controls. The complementary results for sets $\mathcal{S}_1$ and $\mathcal{S}_2$ are presented in \cref{app:CX_PINN_robustness}.

On the other hand, in addition to achieving high fidelity, the control functions predicted by our PINN are also represented by smooth curves, do not present rapid oscillations and satisfy the additional condition $\frac{\partial u}{\partial t}(0)=\frac{\partial u}{\partial t}(T)=0$. This ensures smooth switching at the initial and final times. These features complement gate-design approaches such as those in Refs.~\cite{kuznetsov2025modellingoptimalgenerationarbitrary,c76s-d7pw,Lauten2025} by combining high fidelity with explicit pulse boundary conditions and independent dynamical validation.

\begin{figure}[tbp]
    \centering
    \begin{overpic}[width=\linewidth]{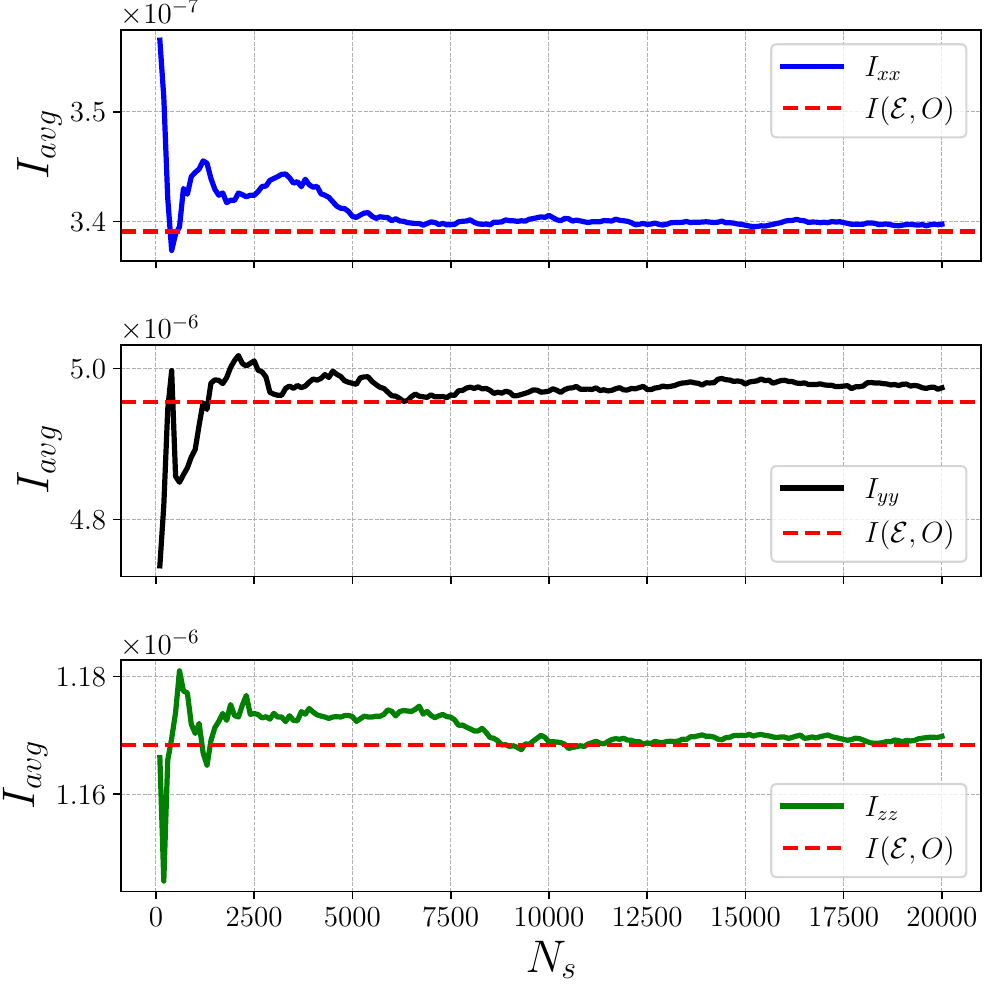}
        \put(30,100){(a) $\sigma_x\otimes\sigma_x$ coupling}
        \put(30,68){(b) $\sigma_y\otimes\sigma_y$ coupling}
        \put(30,36){(c) $\sigma_z\otimes\sigma_z$ coupling}
    \end{overpic}
    \caption{Average infidelity convergence for the CX gate. Average infidelity estimated from \cref{eq:Fm} as $N_s$ increases for the different coupling configurations. (a) $\sigma_x\otimes\sigma_x$, (b) $\sigma_y\otimes\sigma_y$, and (c) $\sigma_z\otimes\sigma_z$. The red dashed line represents the average gate infidelity $I(\mathcal{E},O)=1-F_{\mathrm{avg}}(\mathcal{E},O)$ (\cref{eq:Nielsen}): (a) $I(\mathcal{E},O)=3.39\times10^{-7}$, (b) $I(\mathcal{E},O)=4.96\times10^{-6}$ and (c) $I(\mathcal{E},O)=1.17\times10^{-6}$.}
    \label{fig:approach_CNOT_set3}
\end{figure}

The high fidelity performance of the CX gate implementation is represented in terms of the infidelity behavior in \cref{fig:approach_CNOT_set3}. The average infidelity obtained from \cref{eq:Fm} converges to the reference value from Eq.~\eqref{eq:Nielsen}, as the number of randomly sampled pure states increases. This convergence holds for the three qubit-qubit couplings as indicated by the red dashed line.

\begin{figure}[tbp]
    \centering
    \includegraphics[width=\linewidth]{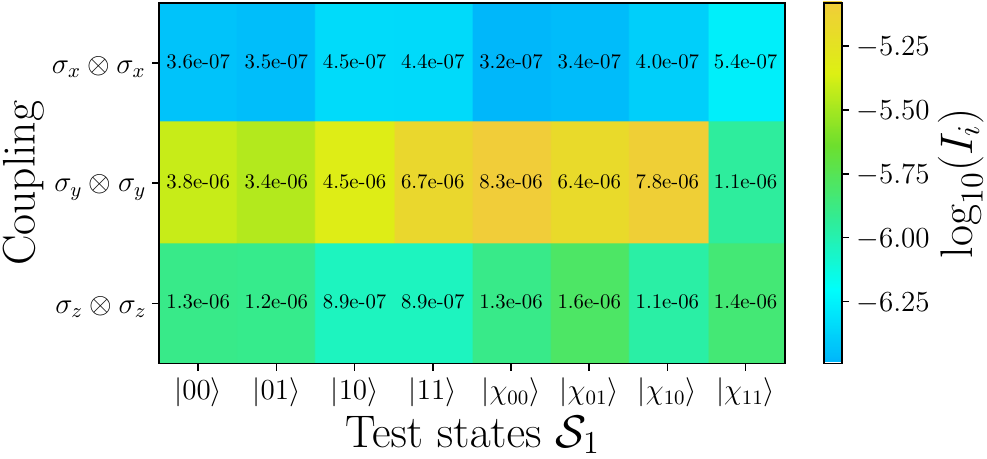}
    \caption{Testing of the PINN-predicted control functions for each coupling. Infidelity heat map computed on the set of states $\mathcal{S}_1$ for each coupling. The colored bar shows the infidelity exponent $p=\log_{10}(I_i)$. The corresponding average infidelities are $\sim4\times10^{-7}$, $\sim5.24\times10^{-6}$, and $\sim1.2\times10^{-6}$ for $\sigma_x\otimes\sigma_x$, $\sigma_y\otimes\sigma_y$, and $\sigma_z\otimes\sigma_z$, respectively.}
    \label{fig:heatmap_set3}
\end{figure}

Furthermore, to confirm these robust results, we also evaluate the predicted control functions on a \textit{test} set of states outside the PINN training. Specifically, set $\mathcal{S}_3$ is used for training, while set $\mathcal{S}_1$, which is a set of mutually unbiased bases different from $\mathcal{S}_3$, is used for testing. Hence, the tested infidelity can be seen as a heat map  in \cref{fig:heatmap_set3} for the eight states of $\mathcal{S}_1$. These results evidence that the predicted control functions reproduce the desired CX gate for states that were not included in the training set with high fidelity (every box in \cref{fig:heatmap_set3} shows the corresponding infidelity value for each coupling and each considered basis state). The $\sigma_y\otimes\sigma_y$ solution exhibits larger test infidelities than the other two coupling configurations. This corroborates the above analysis on the behavior of the control functions shown in \cref{fig:CF_CNOT_S3}. Finally, our results also show that considering control functions along the Pauli axes where the qubit-qubit coupling takes place helps to improve the fidelity of the objective gate. A reduced-control comparison for the $ZZ$ coupling is presented in Appendix~\ref{app:CX_PINN_robustness}. A discussion on the Monte Carlo distribution of the gate infidelity, the fidelity behavior under noisy channels, as well as the dependence on the number of control functions, is also presented in \cref{app:CX_PINN_robustness}.

\subsection{High-fidelity CX gate for the spin-like XY coupling}

Spin-like XY interaction, which can arise from dipole-mediated coupling, is a natural mechanism for exchanging excitations between two two-level systems such as superconducting transmons or phase qubits \cite{majer2007}, polar molecules \cite{yan2013} and Rydberg atoms with opposite parity states \cite{barrero2015}. This interaction, described by a $(\sigma_x\otimes\sigma_x + \sigma_y\otimes\sigma_y)$ coupling, produces population exchanges between the intermediate states $\ket{01}$ and $\ket{10}$, which become important for SWAP-like gates and entanglement generation \cite{abrams2019implementation,susa2018}. Thus, in this section we focus on the PINN-predicted quantum control functions for the CX gate in a two-qubit system with the reference XY interaction Hamiltonian \cite{Chen2025}:

\begin{align}
    H_3(t)&=\dfrac{\Delta}{2} (\sigma_z \otimes \mathds{1} +\mathds{1}\otimes  \sigma_z) + \dfrac{g}{2}(\sigma_x\otimes\sigma_x + \sigma_y\otimes\sigma_y) \nonumber \\
    &+ \dfrac{\Omega_{1}}{2} \sigma_x \otimes \mathds{1} + \dfrac{\Omega_2}{2} \mathds{1}\otimes  \sigma_x , \label{eq:XY_hamiltonian}
\end{align}

where $\Delta$ denotes the detuning of both qubits with respect to the external excitation, $g$ is the XY coupling strength and $\Omega_i$ stands for the strength of the local external excitation field acting on the $i$-th qubit.

Following the discussion in previous sections, we implement the CX gate by considering five different configurations of the reference Hamiltonian as follows (local controls along $\sigma_x$ are considered to be time-dependent such that $\Omega_i = u_x^{(i)}(t)$): (i) Detuning $\Delta$ is considered to be modulated in time and distinct for each qubit, i.e., $\Delta_i = u_z^{(i)}(t)$. (ii) Additional local time-dependent controls along $\sigma_y$ are included (six controls in total as in \cref{sec:PINN_tqCX}). The other three setups consider controls $u_x^{(i)}(t)$ and $u_y^{(i)}(t)$, but fixed detuning on $\sigma_z$ with values (iii) $\Delta/g = 0$, (iv) $\Delta/g = 1/2$ and (v) $\Delta/g = 2$.

\begin{figure}[tbp]
    \centering

    \begin{overpic}[width=\columnwidth]{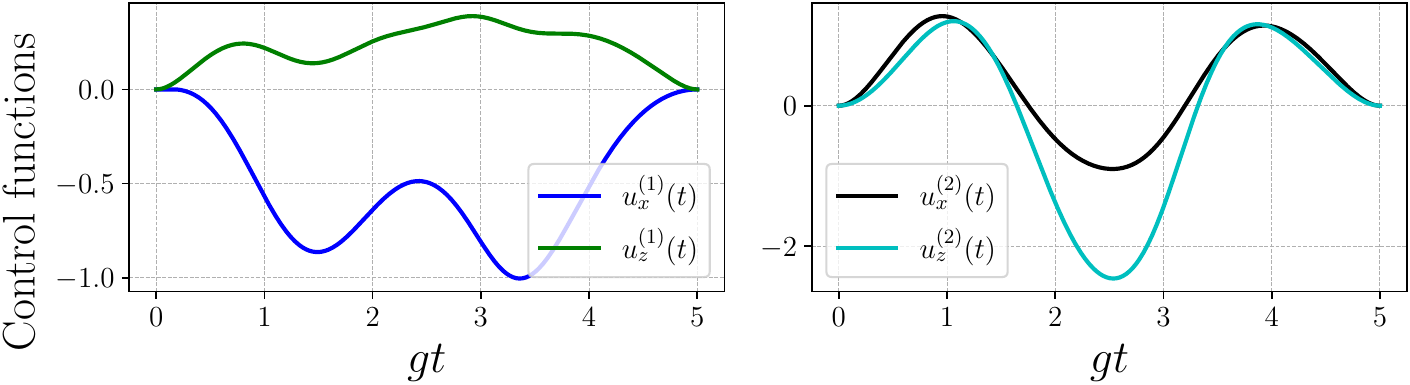}
        \put(0.525\columnwidth,27){\small\text{(i)}}
    \end{overpic}

    \vspace{0.1cm}

    \begin{overpic}[width=\columnwidth]{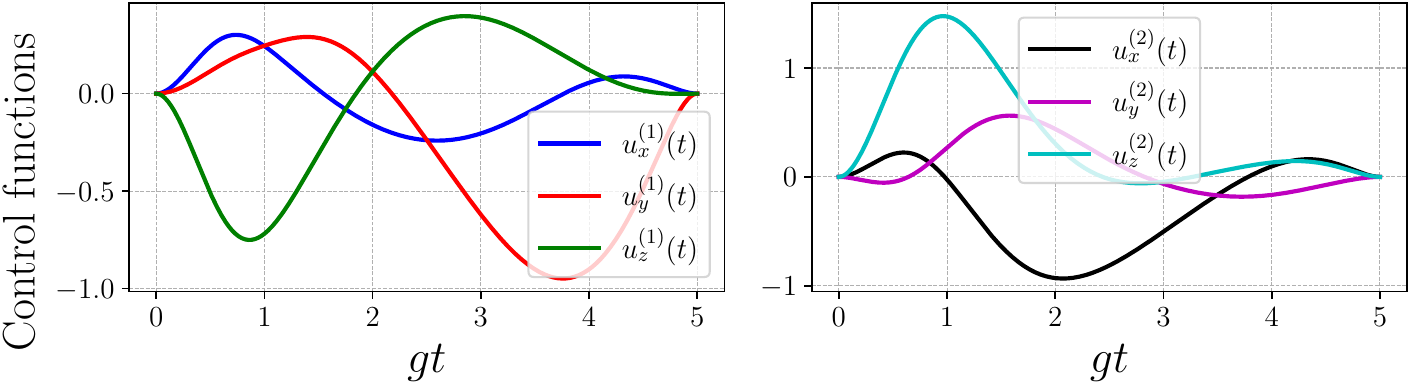}
        \put(0.52\columnwidth,27){\small\text{(ii)}}
    \end{overpic}

    \vspace{0.1cm}

    \begin{overpic}[width=\columnwidth]{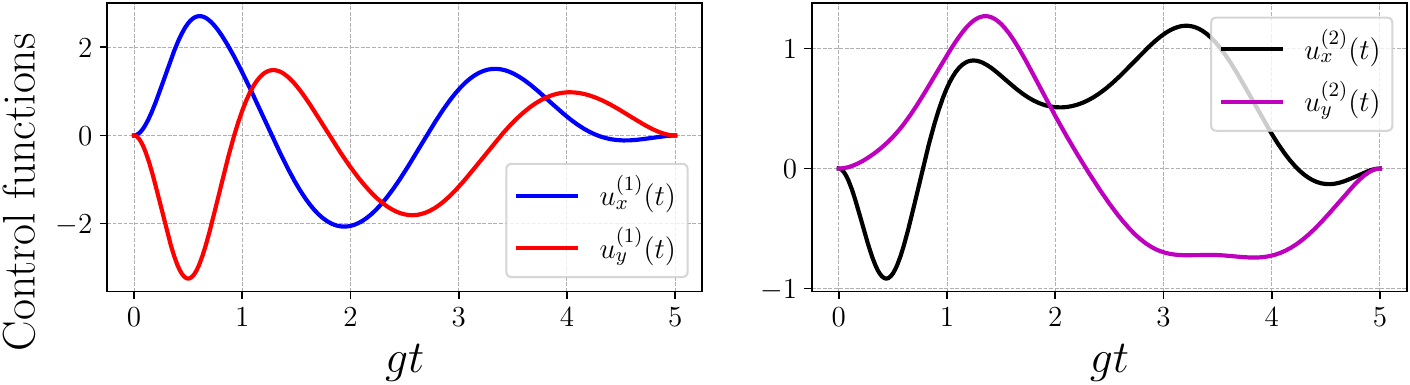}
        \put(0.51\columnwidth,27){\small\text{(iii) }}
    \end{overpic}

        \vspace{0.1cm}

    \begin{overpic}[width=\columnwidth]{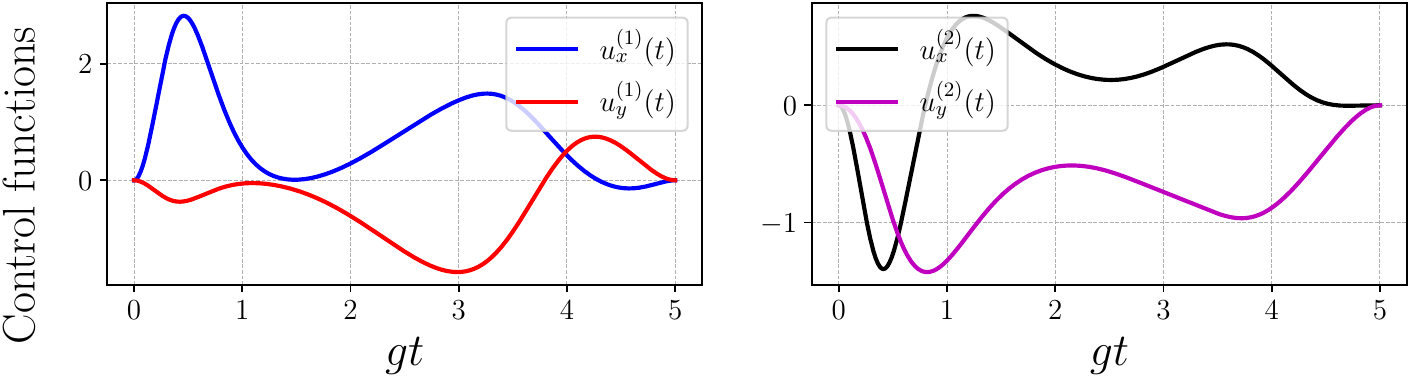}
        \put(0.51\columnwidth,27){\small\text{(iv) }}
    \end{overpic}

        \vspace{0.1cm}

    \begin{overpic}[width=\columnwidth]{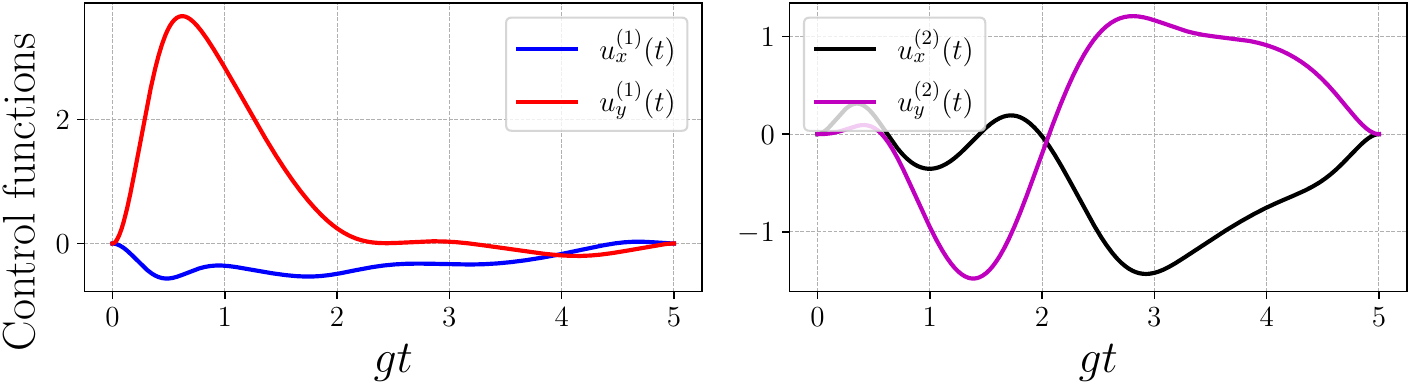}
        \put(0.51\columnwidth,27){\small\text{(v) }}
    \end{overpic}

    \caption{CX gate PINN-predicted control functions for the spin-like XY Hamiltonian  \cref{eq:XY_hamiltonian}. (i) Independent detuning controls, (ii) independent detuning controls with additional $u_y^{(1)}$ and $u_y^{(2)}$ controls. (iii), (iv) and (v) correspond to local controls $u_x^{(1)}$, $u_x^{(2)}$, $u_y^{(1)}$, and $u_y^{(2)}$ for fixed detuning $\Delta/g=0$, $\Delta/g=1/2$, and $\Delta/g=2$, respectively.}
    \label{fig:CF_CNOT_NewH}
\end{figure}

\cref{fig:CF_CNOT_NewH} shows the control functions predicted by the PINN for these configurations. Although the target operation is the same in all cases, the resulting control functions differ notably in both amplitude and shape. In all cases, the control functions predicted by our PINN successfully implement the CX gate, with setup (ii) being the best scenario, reaching an infidelity of $\sim 1.24\times10^{-6}$. The remaining four cases yield slightly larger infidelities: $\sim 2.50\times10^{-6}$ for setup (i), $\sim 3.84\times10^{-6}$ for setup (iii), $\sim 7.94\times10^{-6}$ for setup (iv), and $\sim 3.61\times10^{-5}$ for setup (v). Although these values are higher than that of setup (ii), they remain small, so all five scenarios achieve high-fidelity gates.

From an experimental perspective, the amplitude of each control function corresponds to the strength of the external driving field applied to the system. In the Hamiltonian, the control functions act as time-dependent Rabi frequencies, which are proportional to the field amplitude (e.g., the microwave drive voltage in superconducting circuits), while the corresponding power scales quadratically with it. Consequently, control pulses with smaller amplitudes require weaker excitation and can reduce the required drive strength, subject to platform-specific calibration and bandwidth constraints. Our results show that the six-control setup (ii) requires smaller amplitudes: its peak control amplitudes are approximately half, or less, of those of the largest controls in the four-control configurations. This implies weaker external driving fields, which is advantageous from an experimental standpoint.

\begin{figure}[tbp]
    \centering
    \begin{overpic}[width=\linewidth]{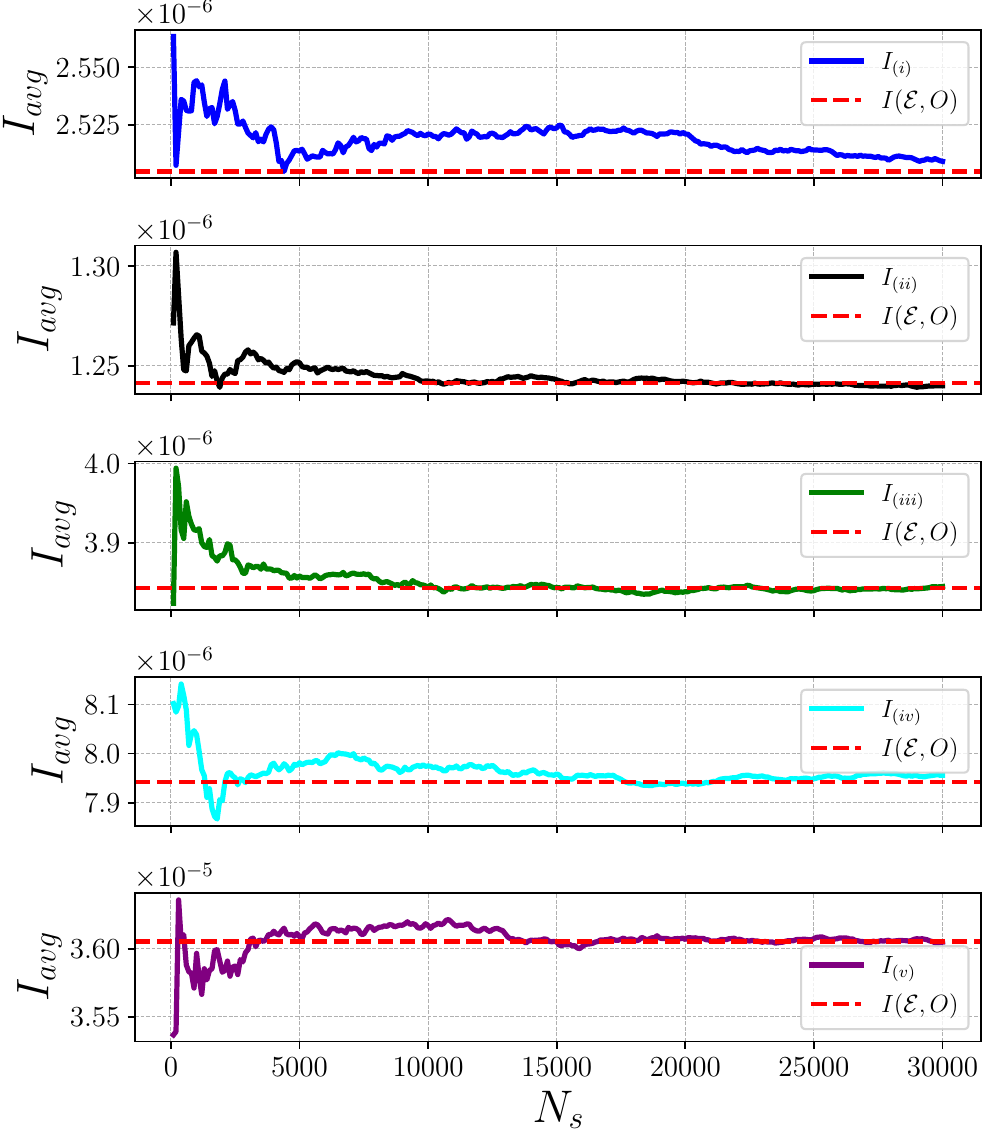}
        \put(0.55\columnwidth,95){(i)}
        \put(0.55\columnwidth,75){(ii) }
        \put(0.55\columnwidth,56){(iii) }
        \put(0.55\columnwidth,37){(iv) }
        \put(0.55\columnwidth,18){(v) }
    \end{overpic}
    \caption{CX gate average infidelity convergence for the XY coupling. Behavior of the average infidelity estimated from \cref{eq:Fm} as $N_s$ increases for the different cases of the XY Hamiltonian. Setups (i)-(v) are the same as those considered in \cref{fig:CF_CNOT_NewH}. The red dashed line stands for the average gate infidelity $I(\mathcal{E},O)$ given through \cref{eq:Nielsen}.}
    \label{fig:approach_CNOT_NewH_set3}
\end{figure}

Although the magnitude of the control functions could be reduced by increasing the regularization strength in the PINN loss function, this also restricts the space of admissible PINN solutions, which typically results in larger infidelities. Hence, a trade-off exists between the control amplitude and the achievable gate fidelity. \cref{fig:approach_CNOT_NewH_set3} shows the convergence of the Monte Carlo estimate of the average infidelity for the five configurations of the XY Hamiltonian. In most cases, the estimate stabilizes around the reference value given by the average gate infidelity of the CX gate, \cref{eq:Nielsen}, as $N_s$ increases. Case (i), however, shows larger fluctuations and a slower stabilization, remaining slightly below the reference value for $N_s=2\times10^4$. Therefore, unlike in the previous sections, we increased the number of sampled states to $N_s=3\times10^4$ in this case to obtain a more stable convergence. These fluctuations concern the finite-sample fidelity estimate. The sampled curves for cases (ii)--(v) stabilize more rapidly in the reported calculations. As before, the robustness with respect to dissipation is left to \cref{app:XY_case}.

\subsection{Control resources for high-fidelity two-qubit gates}

The available local control directions $u_l^{(1,2)}(t)$ are essential to achieve
controllability in the system. In particular, restricting these directions can preserve a
symmetry that imposes an upper bound on the achievable gate fidelity.
For example, consider the $ZZ$ interaction with only local $x$ controls
\begin{align}
H_{zz}(t)
&=J_z\,\sigma_z\otimes\sigma_z \nonumber\\
&\quad+u_x^{(1)}(t)\,\sigma_x\otimes\mathds{1}
+u_x^{(2)}(t)\,\mathds{1}\otimes\sigma_x.
\label{eq:symmetry_H2}
\end{align}
Each term commutes with
$P=\sigma_x\otimes\sigma_x$, so the unitary propagator starting from
$U(0)=\mathds{1}\otimes\mathds{1}$ satisfies $[U(T),P]=0$ for any pulse
shape and duration time $T$. For the fixed target $C=\mathrm{CX}_{1,2}$ in the
computational basis, with the first qubit as the control, define
\begin{equation}
Q\equiv\frac{C+PCP}{2}
=\mathds{1}\otimes\frac{\mathds{1}+\sigma_x}{2}.
\label{eq:symmetry_projected_CX}
\end{equation}
Writing $U=U(T)$, we use cyclicity of the trace, $PUP=U$, and
$C=C^{\dagger}$ to obtain $\operatorname{Tr}(C^{\dagger}U)=\operatorname{Tr}(QU)$. Since $Q$ is a rank-two orthogonal projector and $U$ is unitary,
$|\operatorname{Tr}(QU)|\leq2$. The average gate fidelity over
Haar-distributed pure input states~\cite{nielsen2002simple} therefore
satisfies, for
$\mathcal{E}_U(\rho)=U(T)\rho U^{\dagger}(T)$,
\begin{align}
F_{\mathrm{avg}}(\mathcal{E}_U,C)
&=\frac{|\operatorname{Tr}(C^{\dagger}U)|^2+4}{20}
\leq\frac{2}{5},
\label{eq:symmetry_fidelity_ceiling}
\\
I(\mathcal{E}_U,C)&\equiv1-F_{\mathrm{avg}}(\mathcal{E}_U,C)
\geq\frac{3}{5}=0.6.
\label{eq:symmetry_infidelity_floor}
\end{align}

This bound is independent of the optimization method and applies to unitary evolution under Eq.~\eqref{eq:symmetry_H2}, without additional local rotations before or after the implemented propagator. We remark that this 0.6 value is reported in \cref{fig:2CF_CNOT_S3} (see Appendix~\ref{app:CX_PINN_robustness}). Local fields
along $y$ break the conserved symmetry and remove this particular
obstruction; achieving high fidelity also depends on the remaining
control constraints. The comparison of two- and four-control
configurations presented in this work illustrates how the available control
directions can improve CX fidelities.

\section{Compact control representations and square-pulse playback}
\label{app:chebyshev_all_controls}

Now, we represent the PINN control functions using a Chebyshev expansion to validate their continuous reconstructions and square-pulse approximations in a low-dimensional parameter space. This separates the high-dimensional training parametrization from the exported waveform representation. In particular, Chebyshev expansions have also been used to tabulate neural control pulses~\cite{GUNGORDU2022}; here we separately quantify representation and playback errors across the full collection of archived protocols, and assess gate fidelity after conversion to square pulses as a step toward device-specific implementation.

The Chebyshev polynomials of the first kind are defined by
\begin{equation}
 T_n(x)=\cos(n\arccos x),\qquad -1\leq x\leq 1,
 \label{eq:cheb_definition}
\end{equation}
with $T_0(x)=1$, $T_1(x)=x$, and the recurrence
$T_{n+1}(x)=2xT_n(x)-T_{n-1}(x)$. For each scalar control
$u_\mu(t)$, we use
\begin{equation}
 u_{\mu,K}(t)
 =
 16s^2(1-s)^2\sum_{n=0}^{K-1}c_{\mu n}T_n(2s-1),
 \qquad s=t/T,
 \label{eq:cheb_ansatz}
\end{equation}
where $K$ is the number of coefficients per scalar control.
Thus, a gate representation with $n_{\mathrm{c}}$ scalar controls uses
$n_{\mathrm{c}}K$ coefficients: at the selected $K_\ast=24$, this gives
24 coefficients for Hadamard, 48 for a two-control protocol, 96 for a
four-control protocol, and 144 for a six-control protocol. This counts waveforms, whereas the PINN also represents state
trajectories. The prefactor and its first derivative vanish at both
endpoints, enforcing $u_{\mu,K}(0)=u_{\mu,K}(T)=0$ and
$\dot u_{\mu,K}(0)=\dot u_{\mu,K}(T)=0$.
The polynomial has degree at most $K+3$, hence 27 at $K_\ast=24$.
Its zero continuation outside the gate interval is $C^1$.
Boundary flatness is imposed on the reconstruction: the endpoint
values of sampled inputs do not determine their endpoint derivatives.

\begin{figure*}[tbp]
 \centering
 \includegraphics[width=0.95\textwidth]{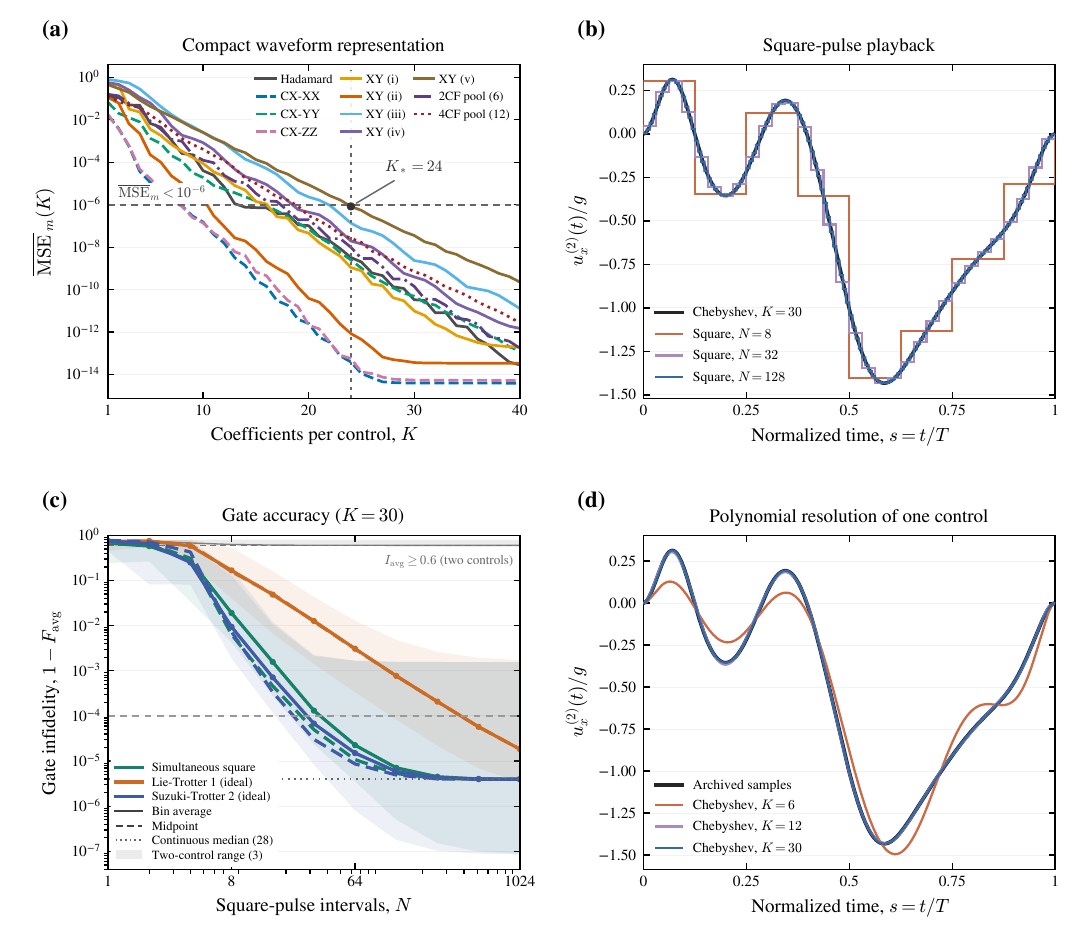}
 \caption{Compact representations and square-pulse playback of 139 learned
scalar controls. (a) Model-averaged MSE,
Eq.~\eqref{eq:cheb_model_mse}, versus $K$ for the $C^1$ basis
$16s^2(1-s)^2T_n(2s-1)$. Eleven families comprise the original nine
models and the pooled two- and four-control cases. Fitted waveforms
satisfy $u(0)=u(T)=\dot u(0)=\dot u(T)=0$ by construction.
The horizontal dashed line marks $10^{-6}$, and the vertical dotted
line marks $K_\ast=24$, the smallest common coefficient count for
which all eleven model-averaged MSEs lie below this threshold.
(b) The $K=30$ fit (black) for $u_x^{(2)}(t)/g$ in XY configuration
(v), $\mathcal S_3$, and midpoint-sampled square pulses with
$N=8,32,128$. (c) Target-gate infidelity for joint square,
first-order Lie--Trotter, and second-order Strang propagation.
Colored curves show medians over Hadamard and the 27 two-qubit
protocols with at least four controls: solid lines use interval
averages and dashed lines use midpoints. Colored bands span the full
minimum--maximum range over those 28 protocols for interval averages;
they indicate protocol variability, not confidence intervals.
The gray envelope and thin solid line show the range and pooled median
of all 18 method/sampling/protocol curves of the three two-control
cases; the horizontal line $I=0.6$ marks their analytical floor.
The gray dotted line is the median continuous-fit target infidelity
for the 28-protocol group. (d) Saved samples of the same control
(black) and fits with $K=6,12,30$.
All propagations use $T=5$ and the corresponding Hamiltonian
normalizations. Ideal square schedules switch abruptly and do not
inherit the polynomial fits' $C^1$ regularity.}
 \label{fig:cheb_models}
\end{figure*}

Each control is sampled on the uniform grid
$t_i=i\Delta t$, with $\Delta t=T/(N_{\mathrm{samp}}-1)$ and
$i=0,\ldots,N_{\mathrm{samp}}-1$, including both endpoints.
We use $N_{\mathrm{samp}}=500$ and $T=5$ in the numerical time units
of each simulation, so $s_i=i/(N_{\mathrm{samp}}-1)$.
For every $K$, the coefficients are fitted independently by solving
the unweighted linear least-squares problem
\begin{equation}
 \widehat{\boldsymbol c}_{\mu,K}
 =
 \operatorname*{arg\,min}_{\boldsymbol c\in\mathbb R^K}
 \frac{1}{N_{\mathrm{samp}}}\sum_{i=0}^{N_{\mathrm{samp}}-1}
 \left[
 u_\mu(t_i)-u_{\mu,K}(t_i;\boldsymbol c)
 \right]^2.
 \label{eq:cheb_ls}
\end{equation}
The design matrix has entries
$\Phi_{in}=16s_i^2(1-s_i)^2T_n(2s_i-1)$, and the least-squares
problem is solved using singular-value decomposition. No
regularization or gate-fidelity refinement is applied. The mean squared error of each fitted control is
\begin{equation}
 \operatorname{MSE}_\mu(K)
 =
 \frac{1}{N_{\mathrm{samp}}}\sum_{i=0}^{N_{\mathrm{samp}}-1}
 \left[
 u_\mu(t_i)-\widehat{u}_{\mu,K}(t_i)
 \right]^2,
 \label{eq:cheb_mse}
\end{equation}
where
$\widehat{u}_{\mu,K}(t)
=u_{\mu,K}(t;\widehat{\boldsymbol c}_{\mu,K})$.
Figure~\ref{fig:cheb_models}(a) reports the arithmetic mean of these
errors for each model $m$,
\begin{equation}
 \overline{\operatorname{MSE}}_m(K)
 =
 \frac{1}{M_m}
 \sum_{\mu\in\mathcal C_m}
 \operatorname{MSE}_\mu(K),
 \label{eq:cheb_model_mse}
\end{equation}
where $\mathcal C_m$ contains the scalar controls associated with
model $m$ and $M_m=|\mathcal C_m|$. The original nine families retain
equal channel weights: $M_m=1$ for Hadamard, $M_m=18$ for each of
CX--$XX$, CX--$YY$, CX--$ZZ$, and dipolar case (ii), and $M_m=12$
for dipolar cases (i), (iii), (iv), and (v). Each two-qubit family
pools its three training sets. Two additional resource families pool
the $XX$, $YY$, and $ZZ$ reduced-control cases trained on $\mathcal S_3$:
$M_m=6$ for two controls and $M_m=12$ for four controls.
The complete analysis therefore comprises 139 scalar controls,
31 gate protocols, and 11 model/resource families.

Every integer $K=1,\ldots,40$ is evaluated. The nested fitting spaces
give nonincreasing least-squares errors in exact arithmetic.
We select the smallest common coefficient count satisfying
$\max_m\overline{\operatorname{MSE}}_m(K)<10^{-6}$ and denote it by
$K_\ast$. This waveform tolerance is motivated by the order of the
representative reported gate infidelities, but it is a separate criterion:
MSE and gate infidelity are different quantities.
The resulting $K_\ast=24$ gives a largest model-averaged MSE of
$8.35\times10^{-7}$, whereas $K=23$ gives $1.66\times10^{-6}$;
dipolar case (v) is limiting in both cases. Thus, $K_\ast$ is optimal
only in the sense of the fewest coefficients meeting the adopted
model-averaged tolerance. Figure~\ref{fig:cheb_models}(a) marks the
threshold and $K_\ast$. These sampled, absolute squared-amplitude
means bound neither individual controls nor errors between samples.

We independently propagate a shape-preserving piecewise-cubic (PCHIP)
interpolation of the saved samples, obtaining $U_{\mathrm{s}}$, and
evaluate the fitted polynomials directly to obtain $U_K$.
This compares representations of the supplied controls without
reevaluating the neural network. Complex double-precision DOP853
propagation uses relative and absolute tolerances $10^{-10}$ and
$10^{-12}$. Tightening both tenfold and halving the maximum step gives
normalized, globally phase-aligned Frobenius differences below
$1.8\times10^{-11}$ for the continuous polynomial propagations.
We distinguish fit-induced infidelity
$1-F_{\mathrm{avg}}(U_K,U_{\mathrm{s}})$ from target infidelity
$1-F_{\mathrm{avg}}(U_K,O)$, using the fidelity between unitary channels.
At fixed $K=30$, the maximum fit-induced infidelity over all 31 protocols
is $1.25\times10^{-10}$, and the largest target-infidelity increase is
$9.16\times10^{-8}$. Thus, waveform fitting and gate preservation
are tested separately rather than inferred from MSE alone.

For the independent playback validation below, we retain the more
conservative $K=30$ for every protocol, whose largest model-averaged
MSE is $3.56\times10^{-8}$, and divide $[0,T]$ into $N$
equal intervals of duration $\delta t=T/N$. Each square-pulse amplitude
is either the value $\widehat u_{\mu,30}$ at the interval midpoint
or its exact average over the interval. Polynomial averages are
evaluated analytically, without reinterpolating the fitted controls.
Writing the resulting frozen Hamiltonian as
$\overline H_j=H_0+\sum_\mu\overline u_{\mu j}H_\mu$, the joint
square-pulse propagator is
\begin{equation}
 U_{\mathrm{sq}}^{(N)}(T)
 =e^{-i\delta t\overline H_{N-1}}\cdots
  e^{-i\delta t\overline H_0}.
 \label{eq:cheb_square_propagator}
\end{equation}
This propagation retains the drift and all local fields simultaneously
within each interval. The square schedule does not inherit $C^1$
regularity: switching between plateaus introduces abrupt changes. Lie--Trotter and symmetric Strang formulas provide
additional approximations by splitting each frozen Hamiltonian into its
Pauli terms in a fixed order. These product formulas are optional:
forming a square waveform does not itself require a Trotter
decomposition. We distinguish the discretization error relative to
the continuous $K=30$ propagator from the additional splitting error
relative to Eq.~\eqref{eq:cheb_square_propagator}.

Figure~\ref{fig:cheb_models}(b) illustrates square-pulse conversion;
panel (d) compares $K=6,12,30$ for the same $u_x^{(2)}/g$ control
in XY configuration (v), $\mathcal S_3$.
For panel (c), resource availability defines two groups before fidelity
is evaluated. The six colored curves show medians for Hadamard and
all 27 two-qubit protocols with at least four controls. The gray
envelope retains all 18 combinations of three methods, two sampling
rules, and the three two-control protocols. The $XX$ case with two local $y$ controls conserves
$P=\sigma_y\otimes\sigma_y$; the $YY$ and $ZZ$ cases with two local
$x$ controls conserve $P=\sigma_x\otimes\sigma_x$. The projected
CX operator has rank two in each case, giving the target-infidelity
floor $I\geq0.6$ in Eq.~\eqref{eq:symmetry_infidelity_floor}.
Their continuous $K=30$ infidelities are approximately $0.797564$,
$0.600354$, and $0.600003$, respectively. The floor does not establish
optimality of these particular pulses.

For $N\in\{1,2,4,8,16,32,64,128,256,512,1024\}$, joint square and
Strang propagation preserve their continuous polynomial references
within infidelity $10^{-5}$ for all 31 protocols from $N=128$ onward
on the tested grid, with either sampling rule. Target infidelity is
$\leq10^{-4}$ for 27 of the 28 resource-group protocols at $N=128$
and all larger tested $N$. The remaining XY configuration (v),
$\mathcal S_2$, has a saved-control infidelity of $1.5603\times10^{-3}$
($1.5604\times10^{-3}$ after fitting), so it remains visible in the
analysis. Its larger target error precedes digital approximation.
These results distinguish accurate playback of a control from the
accuracy of that control's target gate.

Square-pulse tables can be supplied to pulse-programming interfaces such as Qblox/Quantify Scheduler,
QUA, and LabOne Q
\cite{QbloxScheduler2026,QuantifySchedulerPulseAPI2026,QuantumMachinesQUATiming2026,ZurichLabOneQPulseAPI2026}.
However, sampling and scheduling limits differ: Qblox has 1-ns
digital-to-analog converter samples but a 4-ns minimum for separate
real-time operations on a 1-ns grid~\cite{QbloxControlTiming2026}.
QUA distinguishes OPX1000 minima of 4 ns in the configuration and
8 ns for duration overrides from the OPX+ minimum of 16 ns
\cite{QuantumMachinesQUATiming2026}.
At 2.4 GS/s, HDAWG samples are 0.417 ns apart, but independently
played waveforms require 32 samples, with 16-sample increments
(13.33 and 6.67 ns)~\cite{ZurichHDAWGTiming2026}.
These minima do not constrain every plateau inside a longer waveform.
For $\tau=\omega_*t_{\mathrm{phys}}$,
$\delta t_{\mathrm{phys}}=T/(N\omega_*)$ must respect the relevant
grid and minimum when separately scheduled.
Physical claims require frequency/amplitude calibration and finite-bandwidth
validation. Ideal Trotter factors also require generator isolation or
refocusing with an active drift; factor counts are not laboratory times.
For illustration, $\omega_*/(2\pi)=1$ MHz gives
$T_{\mathrm{phys}}\simeq0.796\,\mu\mathrm{s}$ and
$\delta t_{\mathrm{phys}}\simeq6.22$ ns at $N=128$.
At $N=256$, the interval is approximately 3.11 ns, below a 4-ns minimum
for independent operations despite exceeding a 1-ns sample period.
The validated representation preserves the supplied control evolution;
physical waveform delivery and experimental gate accuracy remain
separate device-specific questions.

\section{Conclusions}
\label{sec:conclusions}

In this work, we developed a physics-informed neural network approach to design one- and two-qubit quantum gates using time-dependent control fields. By training simultaneously on initial states drawn from pairs of mutually unbiased bases, the method extends state-to-state control to implementing a target operation on multiple inputs. The network approximates the corresponding state trajectories and control functions while incorporating the quantum dynamics into the training objective. The neural network parametrization enforces the initial conditions and produces pulses with vanishing amplitudes and first derivatives at both initial and final times. Independent propagation, together with agreement between Haar-sampled state fidelities and the average gate fidelity, supports the gate implementations beyond the training states and avoids relying on the training loss as the sole measure of performance.

Under unitary dynamics, the Hadamard benchmark reaches an average gate infidelity of $8.49\times10^{-8}$. For the CX controls trained on $\mathcal{S}_3$, the corresponding infidelities are $3.39\times10^{-7}$, $4.96\times10^{-6}$, and $1.17\times10^{-6}$ for the $XX$, $YY$, and $ZZ$ couplings, respectively. Within the training protocol considered, the interaction also affects how the local fields contribute to the learned solution: the $XX$ and $ZZ$ cases exhibit four dominant control components, whereas the $YY$ solution uses all six. These observations characterize the obtained pulses rather than establish minimum control requirements or fundamental limits on the achievable fidelity. The extension to an exchange-type XY Hamiltonian further illustrates how the same design procedure accommodates different combinations of local controls and fixed or independently controlled detunings.

The open system analysis distinguishes coherent gate accuracy from
sensitivity to dissipation. Reusing the pulses obtained under unitary
dynamics, we quantified the fidelity reduction induced by the specified
dephasing and amplitude-damping channels. In the weak-dissipation
regime, the numerical results reproduce the established first-order
dependence on the dissipative rates~\cite{abad2022universal,abad2025impact}.
For a fixed channel and gate duration, the three CX coupling
configurations exhibit the same leading-order trend, despite their
different control profiles. Thus, a highly accurate coherent
implementation does not by itself imply suppression of dissipative
errors. These results provide a reference for assessing future
optimizations that explicitly include noise in the training objective.

A complementary Chebyshev analysis provides compact representations of 139 scalar controls across 31 protocols. The minimum common coefficient count $K_\ast=24$ brings all eleven model/resource-family mean squared errors below $10^{-6}$, with a largest family mean of $8.35\times10^{-7}$. This criterion applies to family averages. For independent dynamical validation, we retain $K=30$ coefficients per channel, yielding a largest model-averaged MSE of $3.56\times10^{-8}$ and a maximum fit-induced gate infidelity of $1.25\times10^{-10}$. Joint square and Strang propagation preserve the continuous polynomial references within infidelity $10^{-5}$ for all 31 protocols at $N=128$ and all larger tested resolutions, under both sampling rules. For these methods and resolutions, target infidelity is $\leq10^{-4}$ for 27 of the 28 configurations comprising Hadamard and the four-/six-control two-qubit protocols. The remaining XY configuration (v), $\mathcal S_2$, retains its larger baseline infidelity of approximately $1.56\times10^{-3}$. The three two-control cases remain subject to the symmetry-induced target-infidelity floor $I\geq0.6$.

Together, these results extend previous PINN-based gate-design studies~\cite{Lauten2025} by combining gate synthesis through shared controls acting on multiple input states with a systematic characterization of interaction and local-field configurations. The explicit pulse boundary conditions, independently assessed gate accuracies, and quantified Chebyshev approximations connect the learning formulation to the properties of the resulting control waveforms. The dissipative analysis provides a quantitative baseline for distinguishing coherent synthesis errors from noise-induced infidelity. Future work should incorporate explicit amplitude and bandwidth constraints into the optimization and validate pulse implementation using calibrated amplitudes, timing, and transfer functions, allowing pulse complexity, physical constraints, measurement error, stochastic noise, and gate accuracy to be assessed together.

\begin{acknowledgments}
The authors remain responsible for the manuscript's content. C.E.S. thanks Francisco González, Alejandro Gómez and Kevin Joven for fruitful discussions.  L.F. and C.E.S. acknowledge support from the University of Córdoba (Grant No.~FCB-10-25). A.M.  acknowledges support from Universidad Nacional de Colombia, Medellín Campus (Grant Nos. 63067 and 67048). A.N.\ acknowledges support from Fondecyt Regular No.~1251131, Fondecyt Exploraci\'on No.~13250014, and Anillo Tem\'atico ATE 250066. We acknowledge the computational resources provided by the Quantum Ada workstation of the QSTATE group at the Department of Physics, Universidad Técnica Federico Santa María, Chile. 
\end{acknowledgments}

\section*{Data and code availability}
Available from the corresponding authors upon reasonable request.

\appendix

\section{Additional CX gate validation and training analysis}
\label{app:CX_PINN_robustness}

\begin{figure}[tbp]
    \centering
    \includegraphics[width=\linewidth]{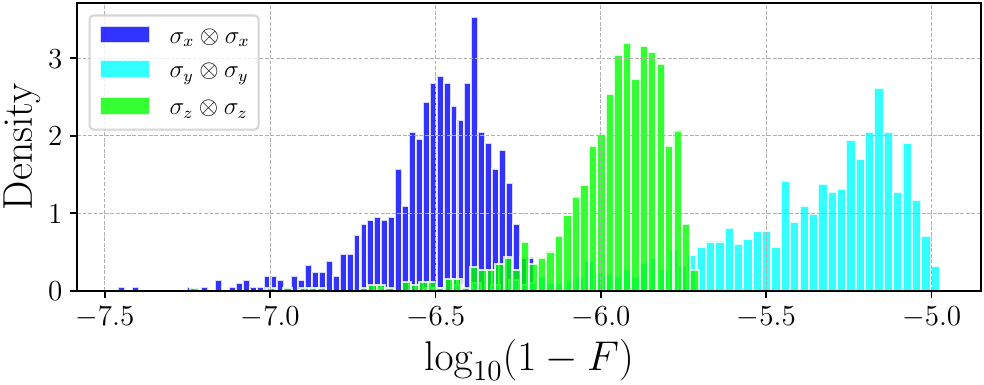}
    \caption{Monte Carlo distribution of CX state infidelities. Distribution of $p=\log_{10}(1-F)$ at the final evolution time $T$, evaluated over $N_s=10^{3}$ Haar-random initial states for the couplings $\sigma_x\otimes\sigma_x$ (blue bars), $\sigma_y\otimes\sigma_y$ (magenta bars), and $\sigma_z\otimes\sigma_z$ (green bars).}
    \label{fig:Infidelity_CNOT}
\end{figure}

To complement \cref{sec:PINN_tqCX}, we present the Monte Carlo distribution of state infidelities for the CX controls shown in \cref{fig:CF_CNOT_S3}. These distributions were computed using $10^{3}$ Haar-random initial states. The controls obtained for the $x$ and $z$ couplings yield lower sampled average infidelities, $3.38\times10^{-7}$ and $1.15\times10^{-6}$, respectively. The $y$ coupling generally produces larger state infidelities, with a sampled average of $4.93\times10^{-6}$ (\cref{fig:Infidelity_CNOT}).

We also examine the CX fidelity under the specified correlated two-qubit channels: pure dephasing with Lindblad operator $L_z=\sigma_z\otimes\sigma_z$, and amplitude damping with operator $L_-=\sigma_-\otimes\sigma_-$. In the weak-dissipation regime ($\gamma_kT \ll 1$), the first-order average gate fidelities for an ideal implementation, according to \cref{eq:first_order_fidelity} \cite{abad2022universal,abad2025impact}, are
\begin{align}
F_z(\mathcal{E}, O)&= 1-\frac{4}{5}\gamma_zT,\nonumber\\
F_-(\mathcal{E}, O)&= 1-\frac{1}{5}\gamma_-T.
\label{eq:Fdissipative_CNOT}
\end{align}

\begin{figure}[tbp]
    \centering
    \vspace{0.25cm}
    \begin{overpic}[width=\columnwidth]{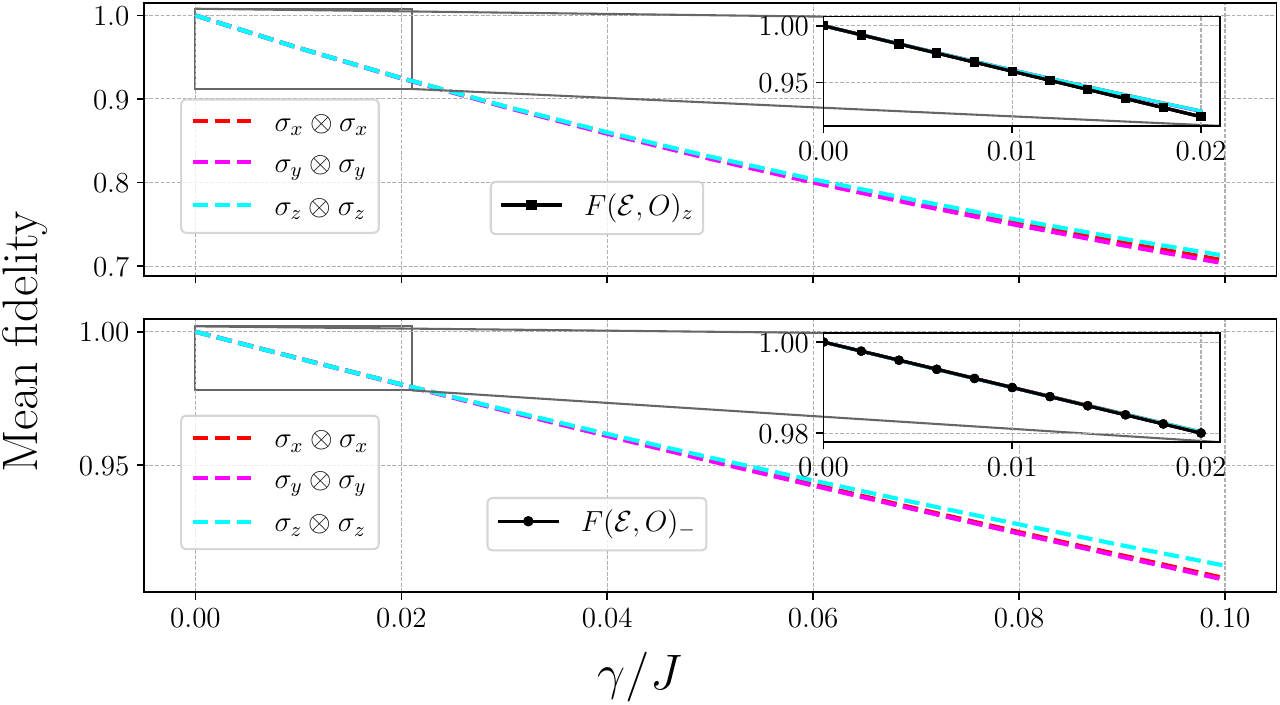}

        \put(12,50){\small\text{$a)$}}
        \put(12,25){\small\text{$b)$}}
    \end{overpic}

    \caption{Average CX fidelity under dissipative channels. Average fidelity evaluated using \cref{eq:Fm} for controls trained under unitary dynamics and propagated with pure dephasing (a) or amplitude damping (b), for the three qubit--qubit couplings. The decay rate varies over $\gamma/J \in [0,10^{-1}]$, where $J=J_x=J_y=J_z$ denotes the coupling strength used in the separate simulations. The insets show the weak-dissipation regime ($\gamma T \ll1$); the dotted black curves represent the first-order fidelities $F_z(\mathcal{E}, O)= 1-\frac{4}{5}\gamma_zT$ and $F_-(\mathcal{E}, O)= 1-\frac{1}{5}\gamma_-T$.}
    \label{fig:Infidelity_PD_AD_CNOT_set3}
\end{figure}

\cref{fig:Infidelity_PD_AD_CNOT_set3} shows the average CX fidelity over $N_s=10^{3}$ random states as a function of the decay rate for the three coupling configurations. Panels (a) and (b) correspond to pure dephasing and amplitude damping, respectively. For each channel, the configurations exhibit similar dependence on the dissipation rate. In the weak-dissipation regime $\gamma T\ll1$ shown in the insets, the numerical results reproduce the linear fidelity reduction predicted by \cref{eq:first_order_fidelity}. For operations confined to the computational subspace, this leading-order contribution depends on the specified dissipative channel and the system dimension, rather than on the implemented unitary operation. The different control profiles therefore do not suppress the leading-order dissipative error. For the operator normalizations and equal rates used here, pure dephasing causes a larger fidelity reduction than amplitude damping, consistent with the respective first-order coefficients. At larger rates, both channels progressively deviate from the linear approximation as higher-order contributions in $\gamma T$ become significant; the deviation is more pronounced for pure dephasing.
\begin{figure*}[tb]
    \centering

    \begin{overpic}[width=0.24\textwidth]{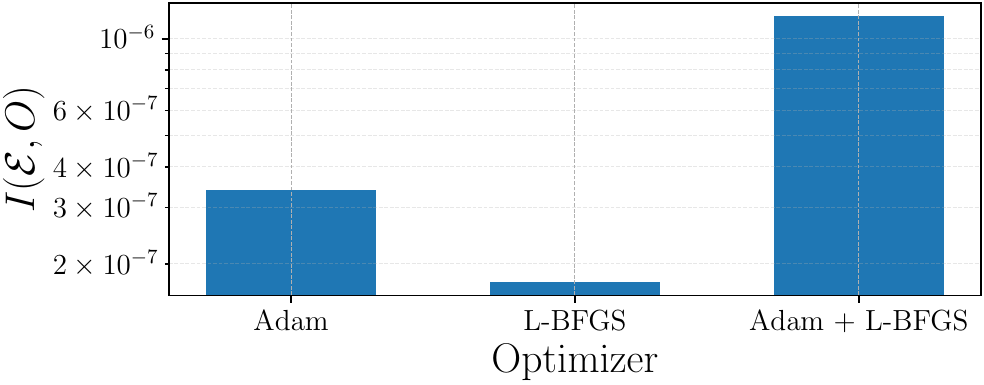}
        \put(12,44){(a)}
    \end{overpic}\hfill
\begin{overpic}[width=0.24\textwidth]{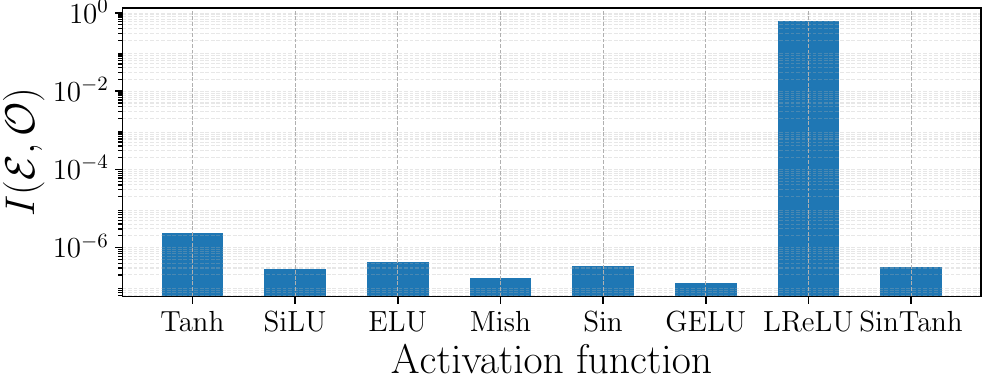}
        \put(12,44){(b)}
    \end{overpic}\hfill
\begin{overpic}[width=0.24\textwidth]{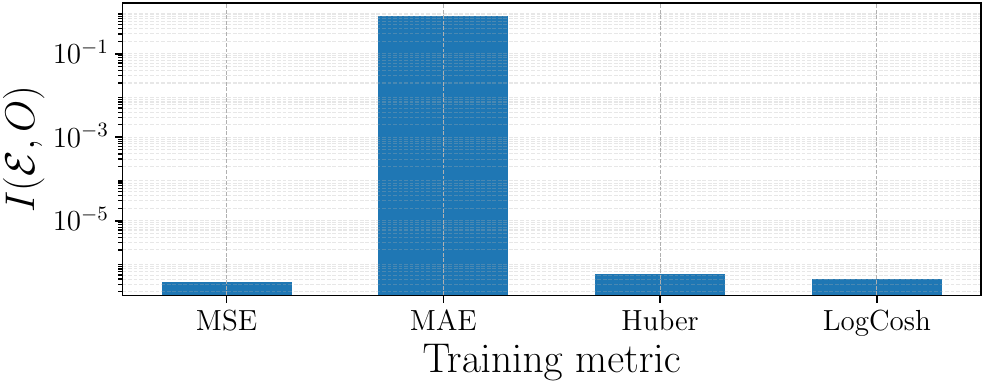}
        \put(12,44){(c)}
    \end{overpic}\hfill
\begin{overpic}[width=0.24\textwidth]{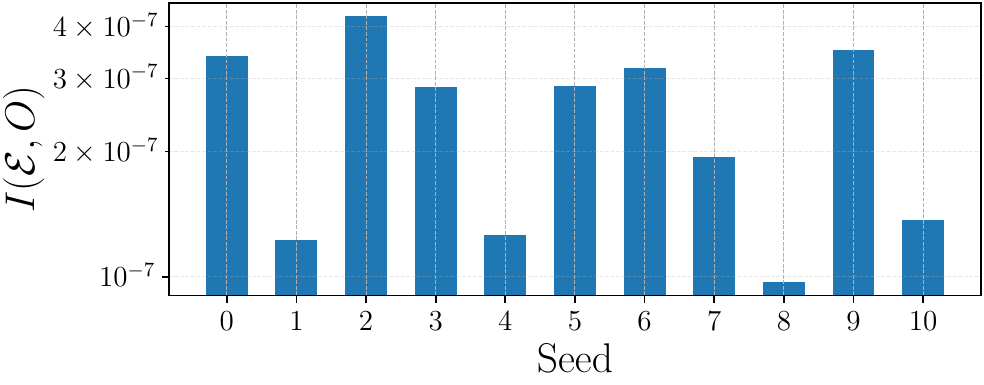}
        \put(12,44){(d)}
    \end{overpic}
\par\vspace{0.2cm}
\begin{overpic}[width=0.24\textwidth]{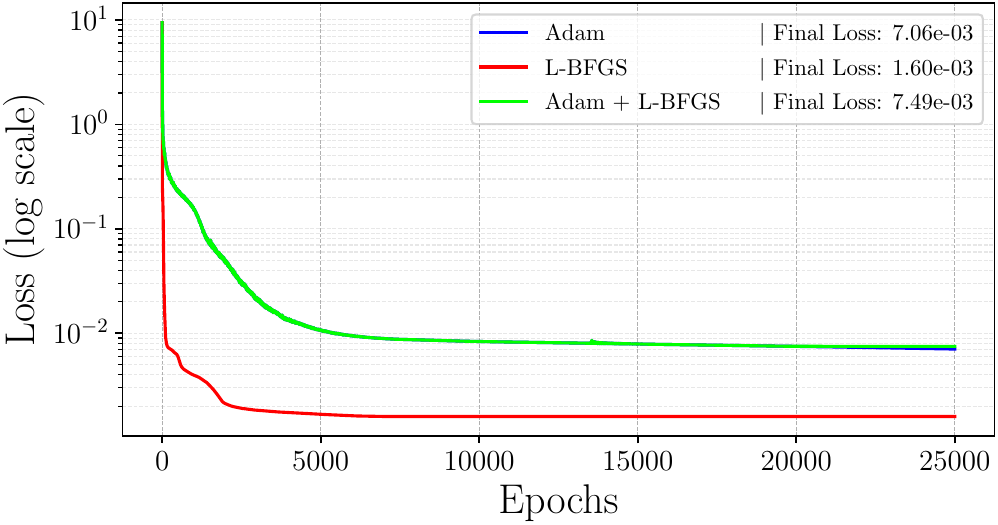}
    \put(12,58){(e)}
    \end{overpic}\hfill
\begin{overpic}[width=0.24\textwidth]{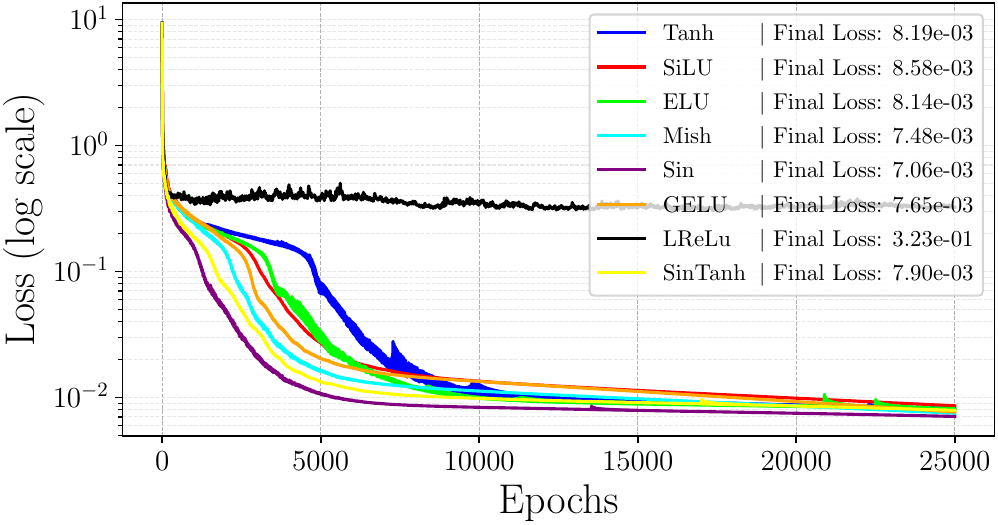}
        \put(12,58){(f)}
    \end{overpic}\hfill
\begin{overpic}[width=0.24\textwidth]{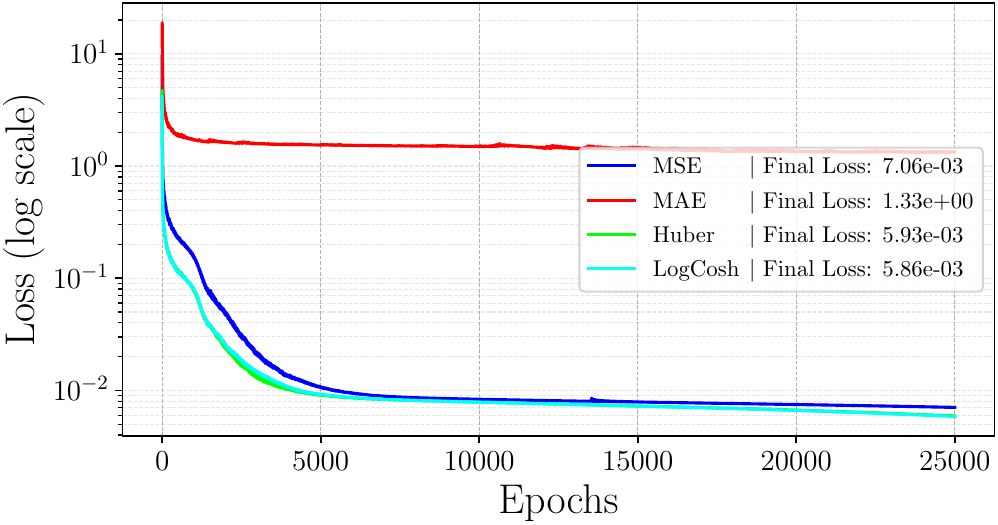}
        \put(12,58){(g)}
    \end{overpic}\hfill
\begin{overpic}[width=0.24\textwidth]{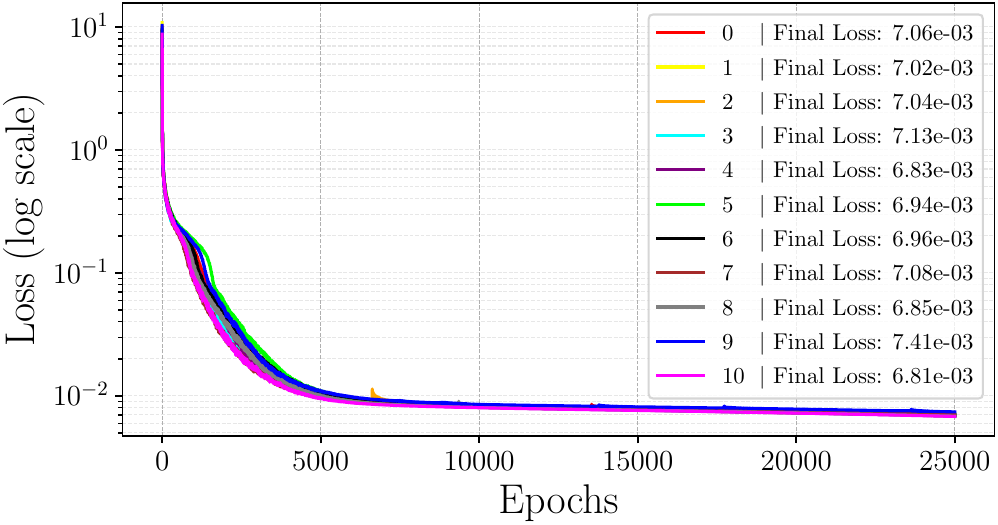}
        \put(12,58){(h)}
    \end{overpic}

    \caption{Sensitivity of the PINN training for the CX gate. Results are shown for the $\sigma_x\otimes\sigma_x$ coupling. Panels (a)--(d) show the average gate infidelity for different optimizers, activation functions, loss metrics, and random seeds, respectively, with the remaining training parameters fixed. Panels (e)--(h) show the corresponding total training losses. The standard deviation across seeds is $1.07\times10^{-7}$.}
    \label{fig:robustness_cnot}
\end{figure*}

As in the Hadamard case, we also carry out the PINN robustness analysis for the CX gate in terms of optimizers, activation functions, loss metrics, and seeds. The same three optimization strategies were compared for CX training. In this case, L-BFGS yields the lowest gate infidelity, while the combined approach, which switches from Adam to L-BFGS after 20,000 epochs, yields the highest infidelity (\cref{fig:robustness_cnot}(a) and (e)).
For the activation functions, SiLU, Mish, and GELU yield similar or lower gate infidelities than Sin (\cref{fig:robustness_cnot}(b)), although Sin retains the lowest training loss (panel (f)). In this case $\alpha=0.56$ and $\beta=0.44$ in the weighted SinTanh function. 
Huber and LogCosh loss metrics exhibit infidelities comparable to those obtained with MSE while MAE shows the worst case for both the infidelity and the loss function (\cref{fig:robustness_cnot}(c) and (g)).
Finally, the multiple-seed training evidenced that more accurate (lower infidelities) scenarios might be achieved by changing the seed as shown in \cref{fig:robustness_cnot}(d), where most of the considered seeds exhibit lower infidelity values than those obtained with seed 0, which was used in the main text, with basically the same loss (panel h).
\begin{figure*}[tb]
    \centering

    \begin{overpic}[width=\columnwidth]{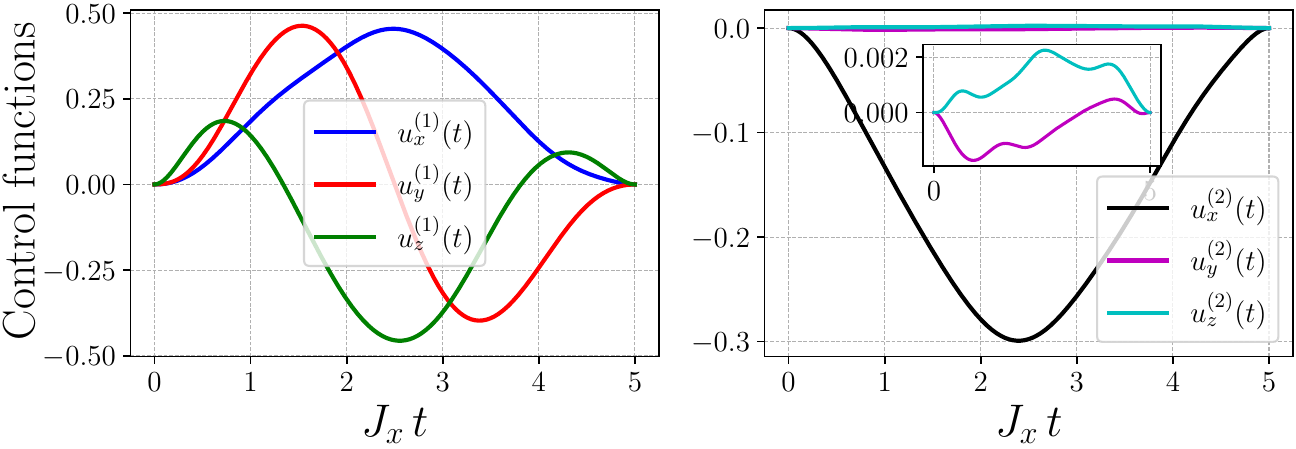}
        \put(10,36){\small\text{(a)} \hspace{23mm} Set of states $\mathcal{S}_1$}
    \end{overpic}\hfill
    \begin{overpic}[width=\columnwidth]{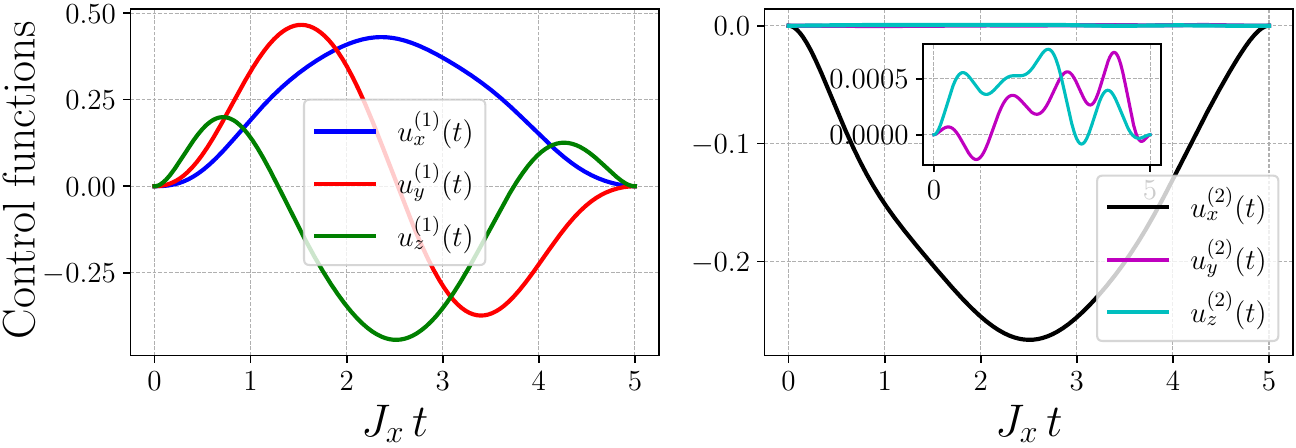}
        \put(10,36){\small\text{(b)} \hspace{23mm} Set of states $\mathcal{S}_2$}
    \end{overpic}

    \vspace{0.40cm}

    \begin{overpic}[width=\columnwidth]{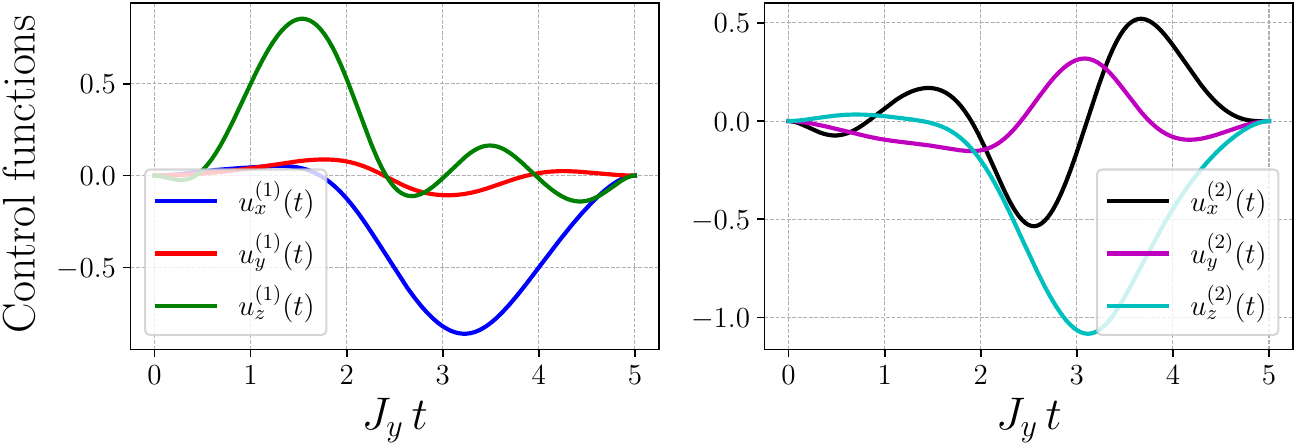}
        \put(10,36){\small\text{(c)}}
    \end{overpic}\hfill
    \begin{overpic}[width=\columnwidth]{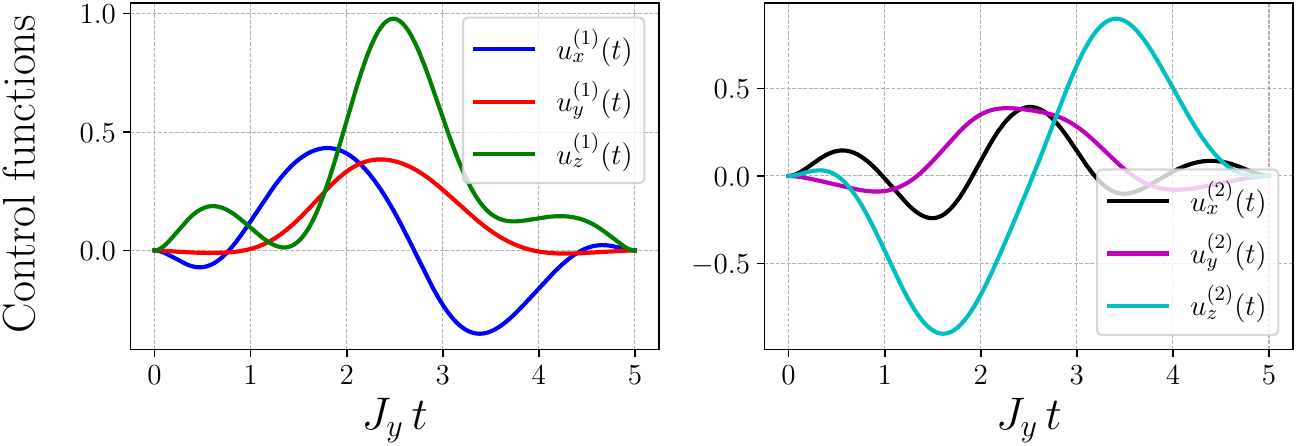}
        \put(10,36){\small\text{(d)}}
    \end{overpic}

    \vspace{0.40cm}

    \begin{overpic}[width=\columnwidth]{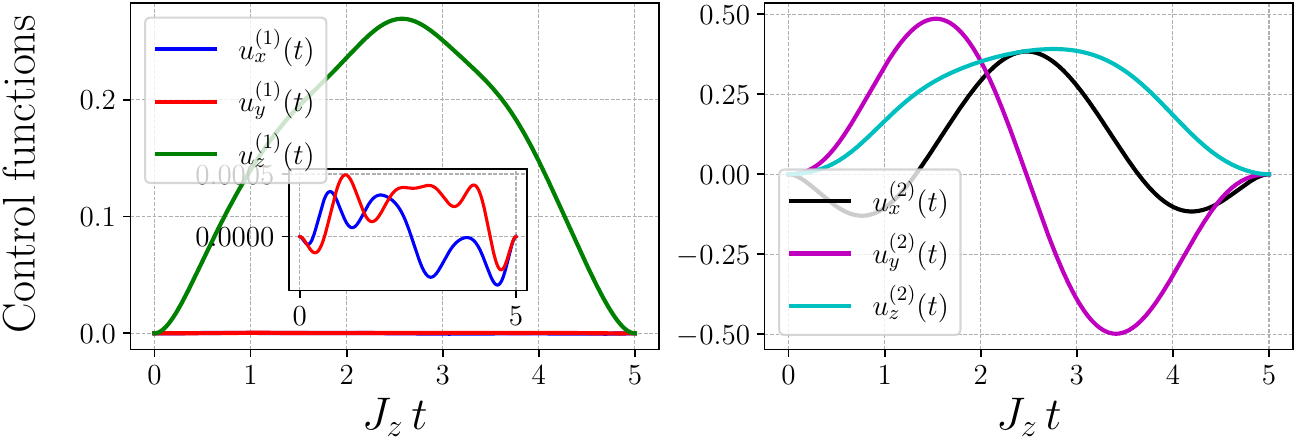}
        \put(10,36){\small\text{(e)}}
    \end{overpic}\hfill
    \begin{overpic}[width=\columnwidth]{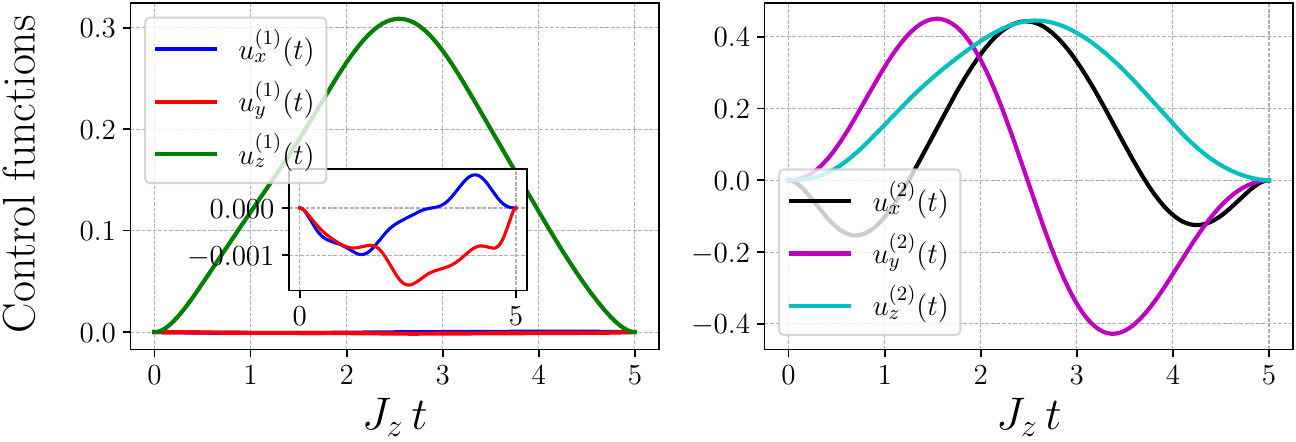}
        \put(10,36){\small\text{(f)}}
    \end{overpic}

    \vspace{0.40cm}

    \begin{overpic}[width=\columnwidth]{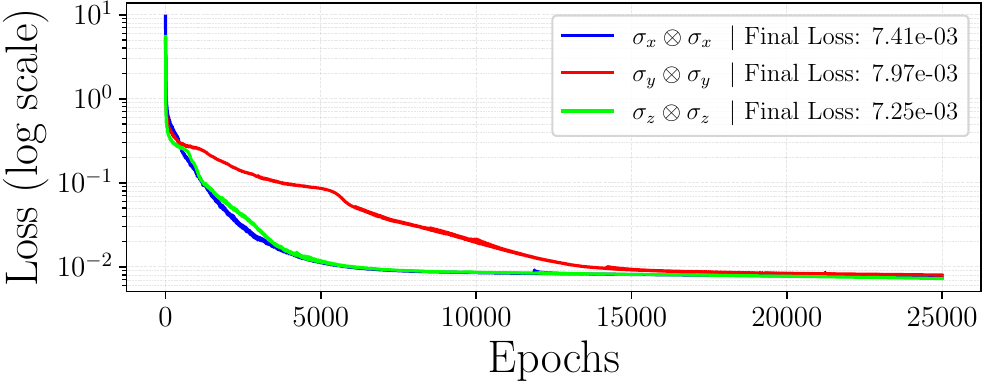}
        \put(15,42){\small\text{(g)}}
    \end{overpic}\hfill
    \begin{overpic}[width=\columnwidth]{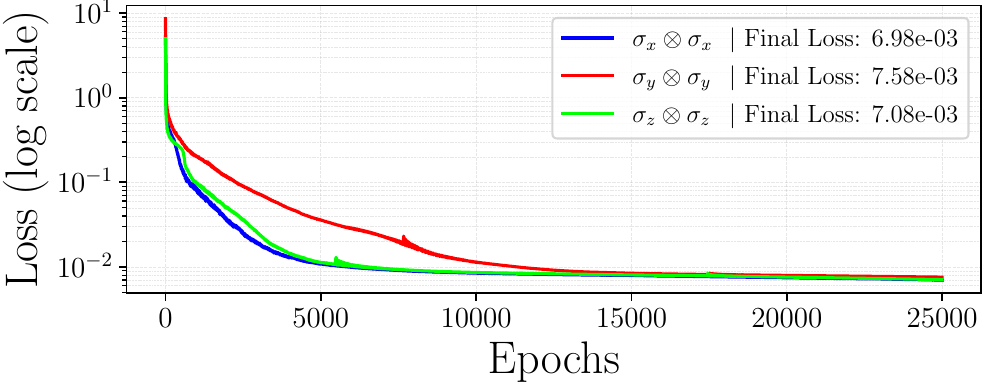}
        \put(15,42){\small\text{(h)}}
    \end{overpic}

    \caption{Control functions and loss for sets $\mathcal{S}_1$ and $\mathcal{S}_2$.  Shape of the control function when sets $\mathcal{S}_1$ (left column) and $\mathcal{S}_2$ (right column) are used in the PINN training. The same coupling setups $\sigma_x\otimes\sigma_x$ (a) and (b), $\sigma_y\otimes\sigma_y$ (c) and (d), $\sigma_z\otimes\sigma_z$ (e) and (f) are considered. (g) and (h) show the corresponding total loss function for each coupling.}
    \label{fig:CF_CNOT_S1S2}
\end{figure*}

Validation on the set of training states. The other two sets of states $\mathcal{S}_1$ and $\mathcal{S}_2$ were also used to carry out additional PINN experiments as a complement to the control functions reported in the main text for the set $\mathcal{S}_3$. \cref{fig:CF_CNOT_S1S2} shows the shape of the control functions in terms of each considered coupling and each set of training states $\mathcal{S}_1$ (left column) and $\mathcal{S}_2$ (right column). Panels (g) and (h) show the corresponding total loss function. Infidelity values similar to those reported in the main text were obtained for these two setups, validating our results. Furthermore, considering $\mathcal{S}_1$ as the training set and taking $\mathcal{S}_3$ as the testing set yields the sampled average state infidelities $I_{\mathrm{avg}}=1.2\times 10^{-6}$ for $\sigma_x\otimes\sigma_x$, $I_{\mathrm{avg}}=9.05\times 10^{-6}$ for $\sigma_y\otimes\sigma_y$, and $I_{\mathrm{avg}}=8.37\times 10^{-7}$ for $\sigma_z\otimes\sigma_z$. All values are comparable with those in the training/testing experiment in \cref{fig:heatmap_set3}, thus confirming the robustness of the PINN-predicted control functions.
\begin{figure}[tb]
    \centering
    \begin{overpic}[width=\columnwidth]{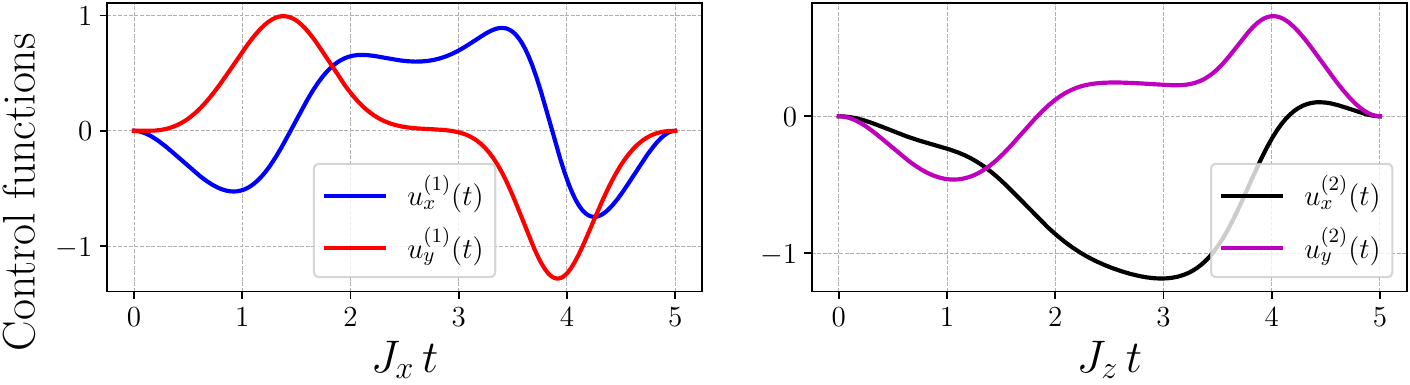}
    \end{overpic}

    \vspace{0.30cm}

    \begin{overpic}[width=\columnwidth]{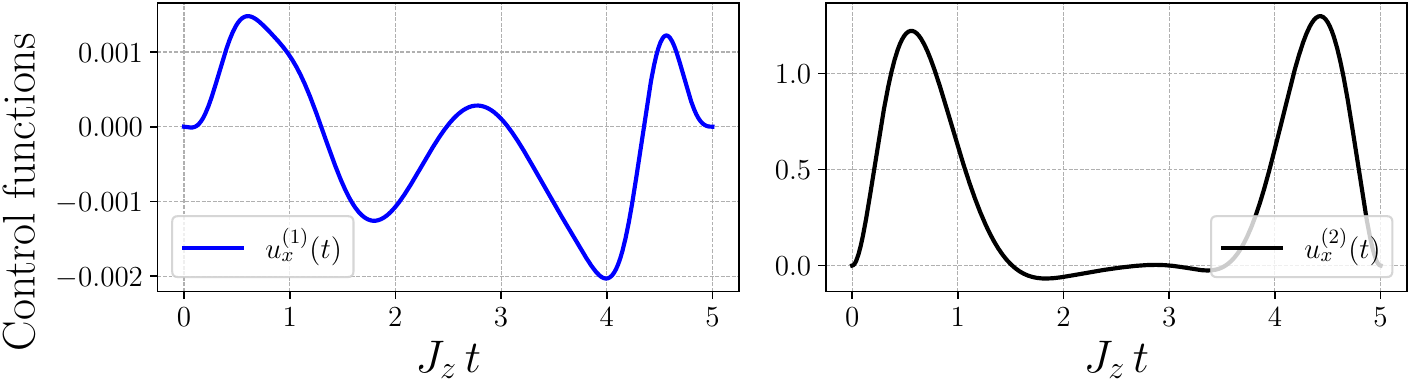}
    \end{overpic}

    \caption{CX gate with a reduced number of control functions. Shape of the control functions for the CX gate by using only four (top) and two (bottom) local controls which are transverse to the considered coupling, here $\sigma_z\otimes\sigma_z$. The average gate infidelities are  $I(\mathcal{E},O)=3.59\times10^{-6}$ and $I(\mathcal{E},O)=6\times10^{-1}$, respectively.}
    \label{fig:2CF_CNOT_S3}

\end{figure}

A natural question that arises from our control function analysis is about the number of controls needed to achieve such a high-fidelity accuracy. To tackle this issue, we run experiments with only four and two control functions which are considered to be transverse with respect to the coupling. \cref{fig:2CF_CNOT_S3} shows the behavior of the control functions when considering four (top) and two (bottom) transversal control functions. It is worth noting that similar high-fidelity values are possible in the former case, with four controls (we got slightly higher infidelity values with respect to the six-control setup). From an experimental viewpoint, this is a reasonable accuracy--resource tradeoff to bear in mind. However, the latter case with only two controls does not achieve good fidelity values.

\section{Sensitivity of Hadamard gate training}
\label{app:H_PINN_robustness}

\begin{figure}[tbp]
    \centering
    \includegraphics[width=\linewidth]{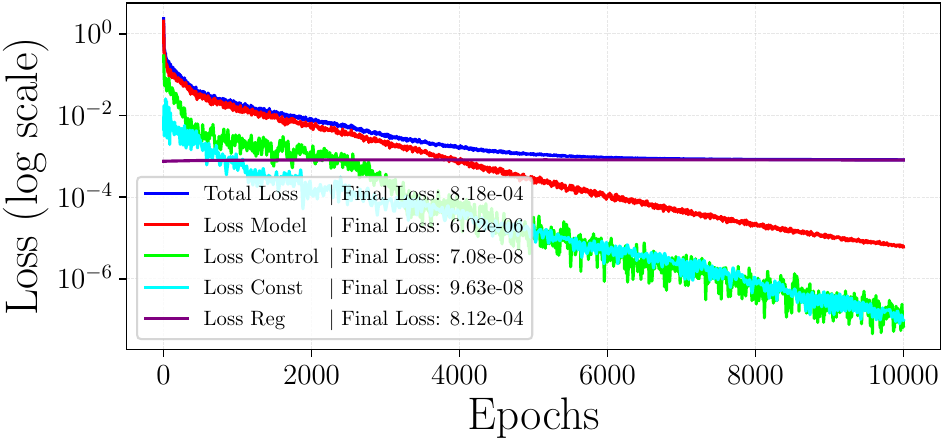}
    \caption{Total loss function and its partial components for the Hadamard gate.}
    \label{fig:partial_loss_Hadamard}
\end{figure}

Here, we present the complete training loss for the Hadamard result shown in the main text and analyze the sensitivity of our PINN architecture to the optimizer, activation function, loss metric, and random seed. Numerical calculations were performed on the Quantum Ada workstation of the QSTATE group at Universidad Técnica Federico Santa María, Chile, equipped with an AMD Ryzen 9 9900X processor, 128 GB of system RAM, and an NVIDIA GeForce RTX 5090 GPU with 32 GB of device memory. The quantum dynamics and gate infidelities were evaluated using QuTiP version $4.7.5$. Both the dynamical evolution and the computation of the unitary propagator $U(T)$ used absolute ($\mathrm{atol}=10^{-13}$) and relative ($\mathrm{rtol}=10^{-12}$) tolerances.

\begin{figure*}[tbp]
    \centering

    \begin{overpic}[width=0.24\textwidth]{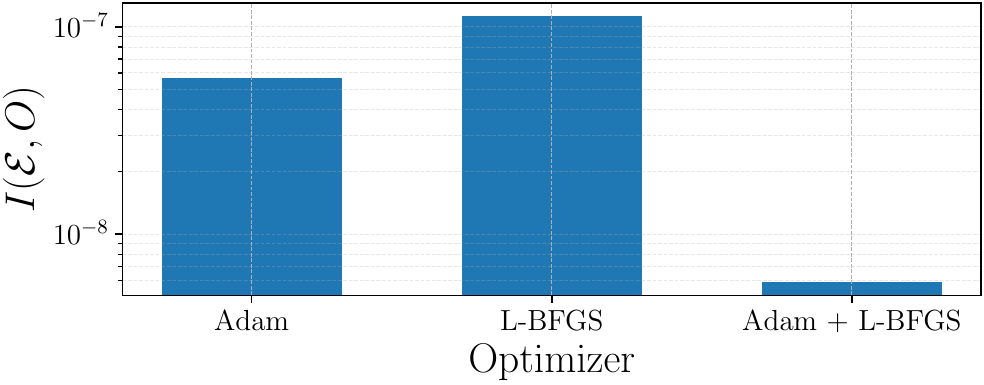}
        \put(12,44){(a)}
    \end{overpic}\hfill
\begin{overpic}[width=0.24\textwidth]{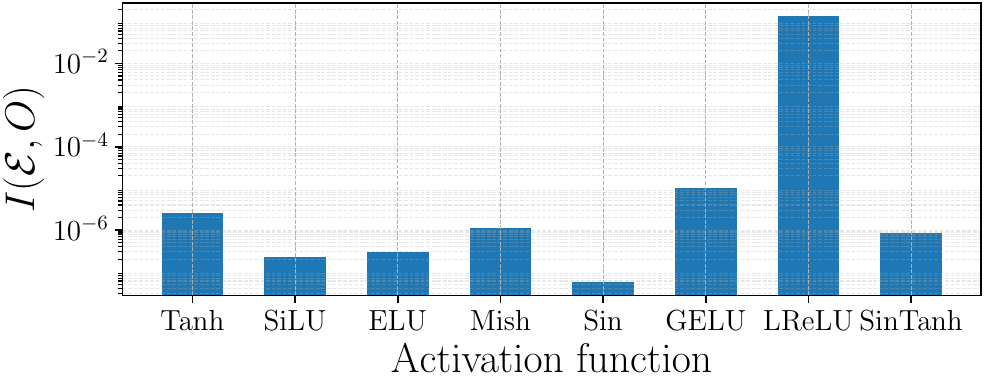}
        \put(12,44){(b)}
    \end{overpic}\hfill
\begin{overpic}[width=0.24\textwidth]{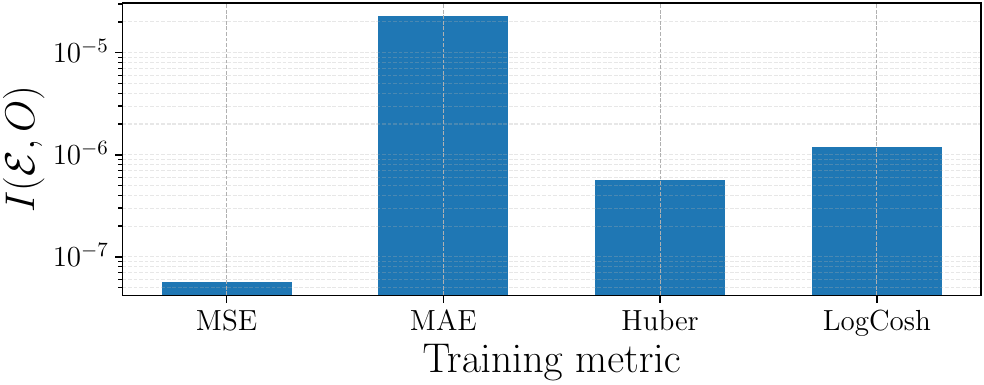}
        \put(12,44){(c)}
    \end{overpic}\hfill
\begin{overpic}[width=0.24\textwidth]{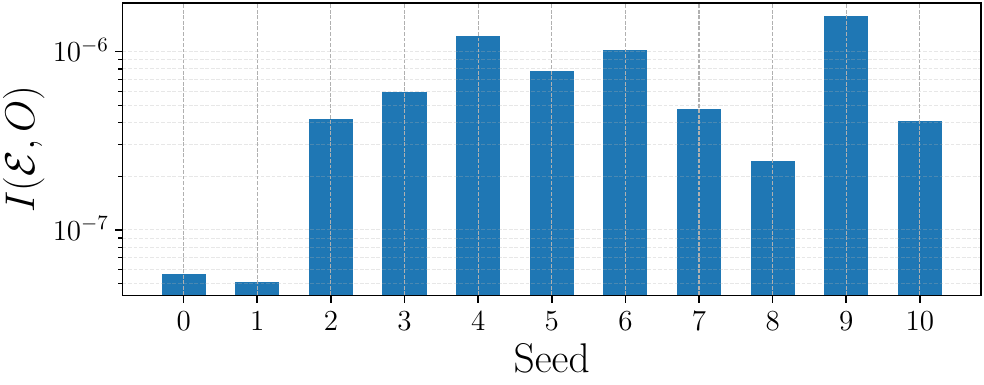}
        \put(12,44){(d)}
    \end{overpic}
\par\vspace{0.2cm}
\begin{overpic}[width=0.24\textwidth]{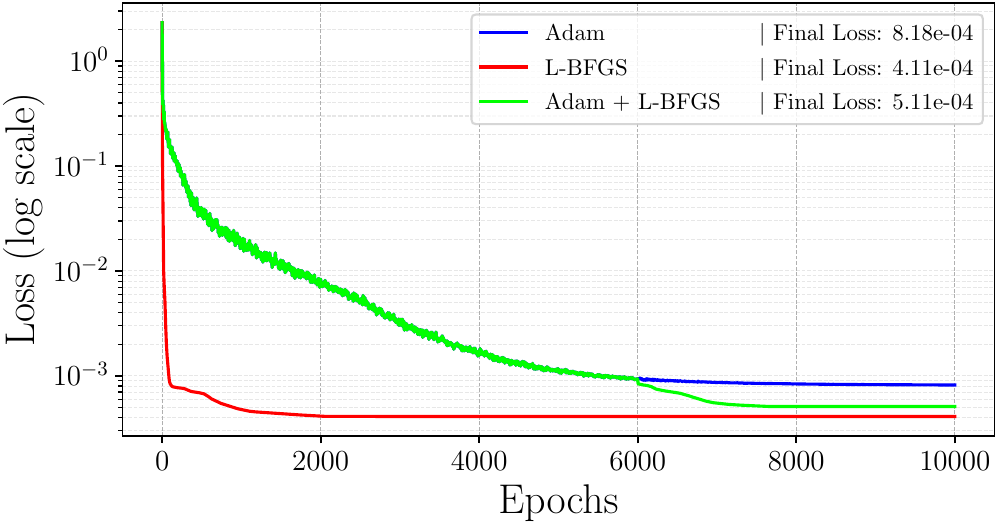}
        \put(12,58){(e)}
    \end{overpic}\hfill
\begin{overpic}[width=0.24\textwidth]{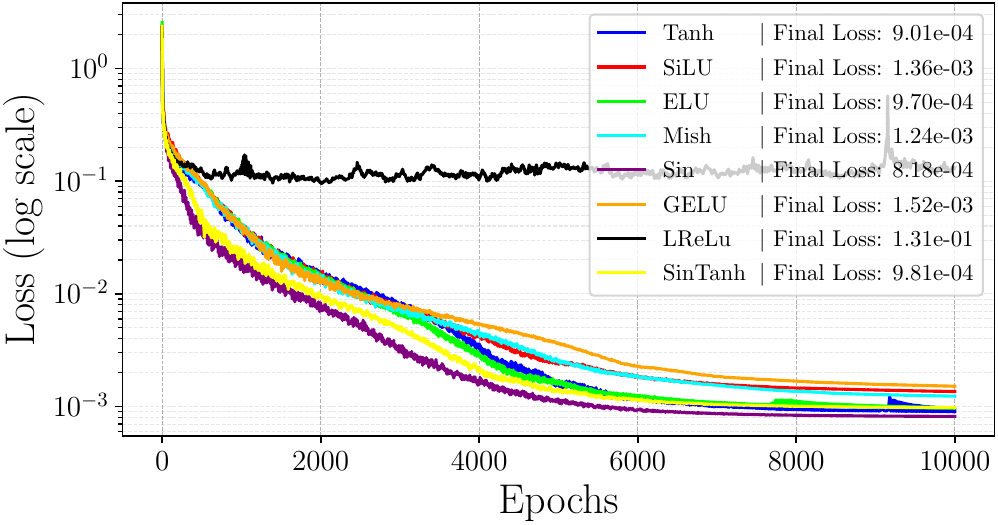}
        \put(12,58){(f)}
    \end{overpic}\hfill
\begin{overpic}[width=0.24\textwidth]{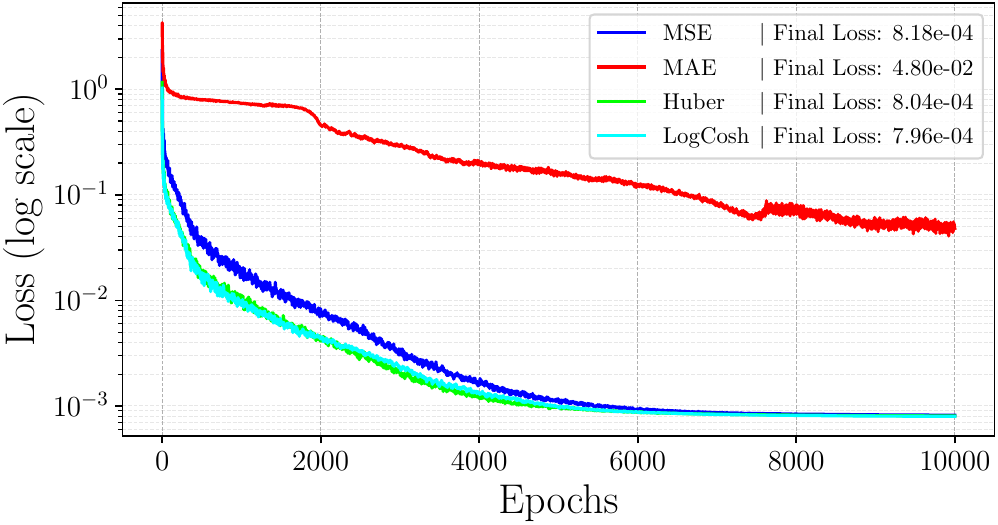}
        \put(12,58){(g)}
    \end{overpic}\hfill
\begin{overpic}[width=0.24\textwidth]{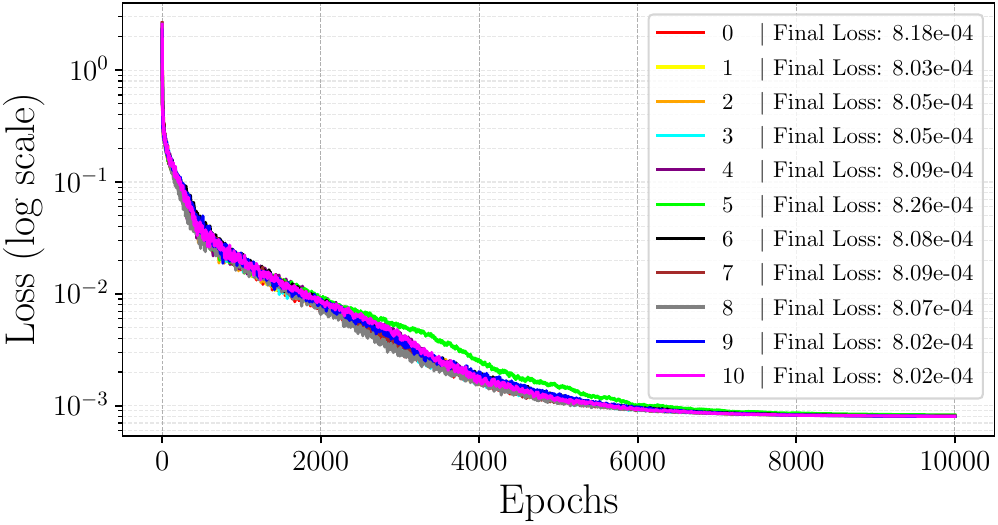}
        \put(12,58){(h)}
    \end{overpic}

    \caption{Sensitivity of PINN training for the Hadamard gate. Final gate infidelity for different (a) optimizers, (b) activation functions, (c) loss metrics, and (d) random seeds. Panels (e)--(h) show the total training loss for the corresponding upper panels.}
    \label{fig:robustness_hadamard}
\end{figure*}

PINN loss function for the Hadamard gate. \cref{fig:partial_loss_Hadamard} shows the total loss and its individual components as functions of the training epoch. The total loss (solid blue line) decreases rapidly during the initial epochs and then slowly approaches a final value dominated by regularization. The losses associated with the physical model (solid red line), the target gate (solid green line), and the density-matrix trace constraint (solid cyan line) decrease throughout training, reaching final values approximately two to four orders of magnitude below the regularization term.

Optimizers. Three optimization strategies were compared: Adam, L-BFGS, and their combination (Adam--L-BFGS), which uses Adam during the first 6000 epochs and L-BFGS for the remaining epochs \cite{rathore2024challenges}. Although Adam--L-BFGS achieves the lowest Hadamard gate infidelity (\cref{fig:robustness_hadamard}(a)), L-BFGS reaches the lowest training loss (panel (e)). The main-text results use Adam, which provides stable training for both the Hadamard and CX gates.

Activation functions. Eight activation functions were compared: Sin, Tanh, SiLU, ELU, Mish, GELU, Leaky ReLU (LReLU), and the weighted combination SinTanh $=\alpha$ Sin $+\beta$ Tanh, where $\alpha$ and $\beta$ are trainable parameters. As shown in \cref{fig:robustness_hadamard}(b) and (f), the Sin activation function used in the main text yields the lowest infidelity and total loss. For SinTanh, the learned weights are $\alpha=0.72$ and $\beta=0.28$.

Loss metrics. In addition to the mean squared error (MSE) used in the main text, we evaluated the mean absolute error (MAE), Huber loss (with $\delta = 1$), and Log-Cosh loss. None of these alternatives yields a lower gate infidelity than MSE, while Huber and Log-Cosh exhibit similar loss convergence (\cref{fig:robustness_hadamard}(c) and (g)).

Seeds. Sensitivity to the initialization of the network parameters was evaluated using the eleven-seed set $\mathcal{S}_{seed}=\{0,1,2,\dots,10\}$. Most runs yield infidelities below $10^{-6}$, with seeds 0 and 1 giving the lowest values (\cref{fig:robustness_hadamard}(d)). The training losses show similar convergence across seeds, as illustrated in panel (h). Seed 0 was used for the main-text results.

\section{Dissipative behavior of the XY coupling PINN average fidelity}
\label{app:XY_case}
\begin{figure}[tb]
    \centering
    \vspace{0.25cm}
    \begin{overpic}[width=\columnwidth]{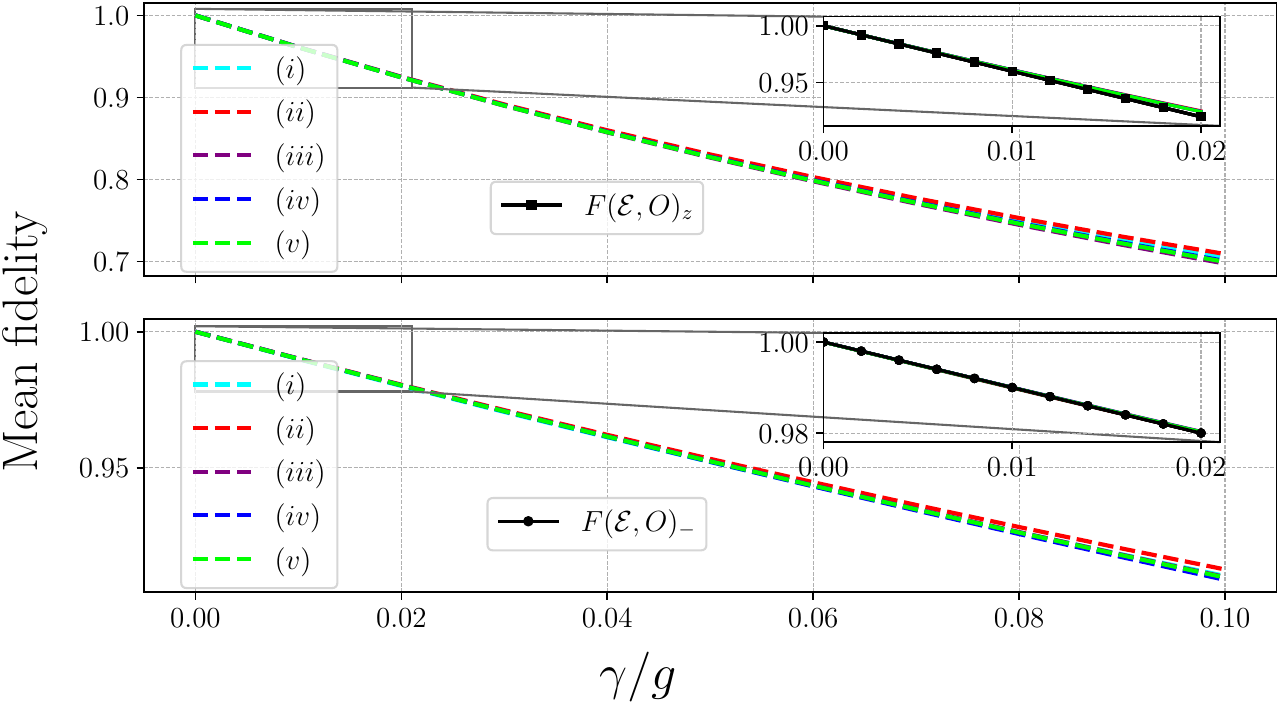}

        \put(12,50){\small\text{$a)$}}
        \put(12,25){\small\text{$b)$}}
    \end{overpic}

    \caption{Average CX fidelity for the XY Hamiltonian configurations (i)--(v). Control functions trained under unitary dynamics are propagated with pure dephasing (a) and amplitude damping (b) as the decay rate varies over $\gamma/g \in [0,0.1]$. The insets show the weak-dissipation regime ($\gamma T \ll1$), with the first-order predictions $F_z(\mathcal{E},O)$ and $F_-(\mathcal{E},O)$ given in \cref{eq:Fdissipative_CNOT}.}
    \label{fig:Infidelity_PD_AD_CNOT_NewH_set3}
\end{figure}

\cref{fig:Infidelity_PD_AD_CNOT_NewH_set3} shows the dissipative behavior of the CX gate under the XY coupling, averaged over $N_s=10^{3}$ random states. Panels (a) and (b) correspond to pure dephasing and amplitude damping for configurations (i)--(v). Because the system dimension and Lindblad operators are the same as in the individual-coupling analysis, \cref{eq:first_order_fidelity} predicts the same leading-order coefficients in the weak-dissipation regime $\gamma T\ll1$. The insets reproduce these linear fidelity reductions. Beyond this regime, the five configurations retain qualitatively similar behavior, with differences becoming more visible as the decay rate increases. Within the rates and channels studied, the distinct control profiles and detunings therefore produce similar dissipative fidelity curves; this observation does not imply that the controls suppress decoherence.

We also repeated the dissipative analysis for both Hadamard and CX using $N_s=2\times10^{4}$ random states at each decay rate. The resulting average fidelity values are basically unchanged with respect to those obtained with $N_s=10^{3}$, both in the weak-dissipation regime and over the full range of rates considered. This agreement indicates stability of the Monte Carlo fidelity estimates with respect to the sample size. 

Additional controls for $\mathcal S_1$ and $\mathcal S_2$ are included in the compact-representation analysis of Sec.~\ref{app:chebyshev_all_controls}. The choice of training set affects gate accuracy: in particular, the larger baseline infidelity of XY configuration (v) trained on $\mathcal S_2$ is reported there.

\clearpage
\bibliography{bibliography,hardware_references}
\end{document}